\documentclass{aa}  

\usepackage{graphicx}
\usepackage{txfonts}
\usepackage{hyperref}
\hypersetup{
    colorlinks=true,
    linkcolor=blue,
    filecolor=blue,      
    urlcolor=blue,
    citecolor=blue,
    pdftitle={Overleaf Example},
    pdfpagemode=FullScreen,
    }
    
\usepackage{xcolor}

\newcommand{\kms}{km s $^{-1}$}
\newcommand{\onep}{\textit{1-pop}\,}
\newcommand{\twop}{\textit{2-pop}\,}
\newcommand{\threep}{\textit{3-pop}\,}

\begin{document}

   \title{Chemo-dynamics of the stellar component of the Sculptor dwarf galaxy I: observed properties}

   \subtitle{}

   \author{
    José María Arroyo-Polonio \inst{1, 2}\thanks{E-mail: jmarroyo@iac.es (IAC)},
    Giuseppina Battaglia\inst{1, 2},
    Guillaume F. Thomas\inst{1, 2}, \\
    Raffaele Pascale\inst{3},
    Eline Tolstoy\inst{4}
    \and Carlo Nipoti\inst{5}.
    \fnmsep
    }

   \institute{Instituto de Astrofísica de Canarias, Calle Vía Láctea s/n E-38206 La Laguna, Santa Cruz de Tenerife, España.    
         \and
             Universidad de La Laguna, Avda. Astrofísico Francisco Sánchez E-38205 La Laguna, Santa Cruz de Tenerife, España.
                     \and
            INAF – Osservatorio di Astrofisica e Scienza dello Spazio di Bologna, Via Piero Gobetti 93/3, 40129 Bologna, Italy
                    \and
            Kapteyn Astronomical Institute, University of Groningen, PO Box 800, 9700AV Groningen, the Netherlands.
        \and
            Dipartimento di Fisica e Astronomia “Augusto Righi” – DIFA, Alma Mater Studiorum – Universita` di Bologna, via Gobetti 93/2, I-40129 Bologna, Italy.
             }

   \date{Submited on \today; received -; accepted -}

% \abstract{}{}{}{}{} 
% 5 {} token are mandatory
 
  \abstract
   {}
   {Recently, both the presence of multiple stellar chemo-kinematic components and rotation in the Sculptor dwarf spheroidal galaxy have been put into question. Therefore, we re-examine the chemo-kinematic properties of this galaxy making use of the best spectroscopic data-set available containing both line-of-sight velocities and metallicities of individual stars.}
   {We carry out a detailed, quantitative analysis on the recent spectroscopic data-set from \cite{Tolstoy2023} that contains high precision velocities and metallicities for 1339 members of Sculptor. In particular, we assess whether Sculptor is best represented by a single stellar population with a negative metallicity gradient or by the super-position of two or more components with different mean metallicity, spatial distribution and kinematic properties. For this analysis, we also include the incompleteness of the spectroscopic data-set.}
    {We find that Sculptor is better described by a two-populations model than by a single-population model with a metallicity gradient. Moreover, given the assumptions of the current modeling, we find evidence of a third population, composed of few stars, that is more extended and metal-poor than the two other populations. This very metal-poor group of stars shows a shift of $\sim$15 km s$^{-1}$ in its average l.o.s. velocity ($v_{los}$) with respect to the rest of the galaxy. We discuss several possible origins for this new population, finding a minor merger as the most likely one. We also find a $v_{los}$ gradient of 4.0$^{+1.5}_{-1.5}$ km s$^{-1}$ deg$^{-1}$ but its statistical evidence is inconclusive and, moreover, its detection is partially driven by the group of stars with off-set velocities.}
   {}


   \authorrunning{Arroyo-Polonio J.-M. et al.}
   \titlerunning{Chemo-dynamics of the Sculptor dSph I: observed properties}
   \maketitle 
%
%-------------------------------------------------------------------

\section{Introduction} \label{Introduction}

The characterization of the internal kinematic properties of the stellar component of galaxies is an important step for unraveling the processes that have shaped their evolution; it is also propaedeutic to the application of dynamical modeling techniques aimed at inferring the galaxy's mass distribution. 

In the Local Group (LG) we have the possibility of carrying out this type of characterization in detail, on a star-by-star basis, for a large sample of dwarf galaxies, reaching down to a regime in stellar mass that is otherwise inaccessible to current facilities. This regime is particularly interesting as it is where the galaxies with the largest dynamical mass-to-light ratios are found \citep[and references therein]{Mateo1998,Walker2011,Simon2019,Battaglia2022b}, making them very valuable probes to test cosmological models. 

For LG galaxies, we can also rely on a wealth of additional, detailed information, such as the stars' chemistry and relative ages, which allow exploring whether there are causal links between observed properties (e.g. does the rotational versus  velocity dispersion support has an impact on the age and metallicity gradients? Do bursts of star formation leave an imprint on the chemo-kinematic properties of the stellar component?). In addition, the impact of environmental effects can be assessed at the level of the individual dwarf galaxies, at least to some extent. For example, from the 3D distance from the large LG spirals, it is possible to identify what dwarf galaxies might have had a negligible level of interaction with these much more massive systems. In particular for Milky Way (MW) satellites, the availability of accurate and precise systemic proper motions derived from Gaia and HST has opened the door to the possibility of relating their observed properties to their orbital history \citep{Sohn2017,Li2021,pace2022,Battaglia2022a}; even though uncertainties in both the satellites' systemic motions and in the mass distribution of the Milky, and how it is assembled over time, imply that such considerations are most robust for the last one or two  pericentric passages \citep[e.g.][]{dsouza2022,Vasiliev2024, Santistevan2024}.

The direction and amplitude of velocity gradients, combined with the information on the galaxy's orbit, can give several useful indications of a galaxy's past life.  As the result of various observing campaigns from multiple groups \citep[e.g.][]{Tolstoy2004, Walker2008}, it has become clear that the stellar component of LG dwarf galaxies is mainly supported by random motions \citep[see e.g.][who find that the measured ratio between ordered and random motions, $V/\sigma < 1$ for 3/4 of the sample]{Wheeler2017}. This is also true for systems that are unlikely to have ever interacted with the large LG spirals, or that might have had experienced one pericentric passage around the MW or M31 at most \citep{Taibi2018, Taibi2022}.
Such findings disfavour, or at least call for a revision of, those models that were suggesting that the pressure-supported dwarf spheroidal galaxies form from tidally-stirred, rotationally supported dwarf galaxies \citep[e.g.][]{Mayer2001,mayer2006, Kazantzidis2011}. On the other hand, the presence of rotation along the minor-axis ({\it prolate rotation}) has been unveiled in And~II \citep{ho2012} and Phoenix \citep{kacharov2017} and linked to merger events \citep{lokas2014,cardonabarrero2021}. In other systems, velocity gradients aligning with the projection of the orbital path on the sky are revealing of tidal stripping (e.g. AntliaII, TucanaIII) \citep{Mutlu2018, Torrealba2019}; and relations have been made between the velocity gradients' amplitude and orbital phase \citep[e.g. approaching vs leaving pericenter][]{Martinezgarcia2023}. 

However, robustly measuring velocity gradients of just a few km/s/kpc entails a significant challenge, depending on factors such as the spatial coverage, velocity uncertainties and sample sizes. Therefore, they are worth revisiting as new data-sets become available, in particular at a time when more precise determinations of the systemic proper motions allow to correct for "perspective" gradients \citep[e.g.][]{kaplinghat2008,Walker2008}.

Another main outcome of the spectroscopic campaigns has been the finding that the stellar component of several LG dwarf galaxies can be described as the superposition of at least two components with different metallicity, spatial distribution and kinematics \citep[dubbed ¨chemo-kinematic components¨ or "multiple chemo-dynamic populations", e.g.][]{Tolstoy2004,Battaglia2006, Walker2011, Fabrizio2016, Kordopatis2016, pace2020}. This has been possible through the combined analysis of line-of-sight velocities and metallicities (or indicators of relative metallicity) for large samples of individual stars. Trying to understand whether these are indeed separate populations can give handles on formation scenarios. For example, multiple populations can form from early mergers (that would puff up a pre-existing old and metal-poor stellar component), followed by later gas accretion, which would form the more metal-rich component \citep[e.g.][]{Benitez-llambay2016}. Or, possibly, these components could be due to star formation bursts, or anyway prominent phases of star formation at given times, rather than a more continuous star formation activity that would favor the formation of metallicity gradients \citep{Revaz2018, Mercado2021}. From a dynamical modeling perspective, the importance of multiple chemo-kinematic populations lies in the fact that they can be used to place tighter constraints on the DM halo density distribution \citep[e.g.][]{Battaglia2008,Walker2011,Amorisco2012,zhu2016,Strigari2017} by relieving the well-known mass-anisotropy-(stellar density) degeneracy \citep{Binney1982}. 

The Sculptor dSph, found at an heliocentric distance of 83.9 $\pm$ 1.5 kpc \citep{Martinezvazquez2015}, was the first LG dwarf galaxy in which the presence of multiple chemo-kinematic components was unveiled \citep{Tolstoy2004} and used as independent tracers of the galaxy's gravitational potential \citep{Battaglia2008}. The rationale for considering these as separate components came mainly from the presence of a well-defined blue- and red-horizontal branch in the color-magnitude diagram of this galaxy, which also exhibited a distinct spatial distribution. Additionally, Schwarzschild modeling of this galaxy naturally led to the identification of two peaks in the energy-angular momentum space, directly related to the two previously known populations  \citep{breddels2014}, giving support to the hypothesis of two distinct stellar components. Dynamical modeling using individually both Sculptor's chemo-kinematic components typically has resulted in a preference for a cored DM halo density distribution \citep{Battaglia2008,Walker2011,Amorisco2012,zhu2016,Strigari2018}, while single component modeling results are more diverse, finding both cores in some cases and cusps in others \citep{Breddels2013, Read2019, Pascale2019, Kaplinghat2019, PascaleThesis}.

\cite{Battaglia2008} also found a l.o.s. velocity gradient of 7.6$^{+3.0}_{-2.2}$ km s$^{-1}$ deg$^{-1}$ along the projected major axis. Nonetheless, the evidence for a significant velocity gradient in this galaxy has weakened, both in studies that have used Gaia DR1+HST or Gaia eDR3 systemic proper motions to correct  the earlier spectroscopic samples for perspective velocity gradients \citep[e.g.][]{Massari2018, martinezgarcia2021} as well as in new larger data-sets with improved, Gaia-aided memberships \citep{Tolstoy2023}. Additionally, in the initial analysis of \cite{Tolstoy2023} the two previous chemo-kinematic components are no longer as easily identifiable as in earlier studies; while on the other hand, the most distinct kinematics is detected for stars with [Fe/H]$<-2.5$, whose average velocity is shifted of a few km/s from that of the rest of the stars. 

Our goal is to perform a comprehensive and quantitative chemo-dynamical analysis of the stellar populations in Sculptor, using the new data-set of \cite{Tolstoy2023} and considering simultaneously the information on the stars's spatial location, l.o.s. velocity and metallicity. We aim to untangle whether Sculptor is better described by a single population with a negative metallicity gradient or whether it is composed of distinct stellar components, and to revisit the presence of a possible $v_{los}$ gradient. The results of this analysis will inform future dynamical modeling (Arroyo-Polonio et al. in prep).

In Sec.~\ref{sec:Data} we will briefly describe the main characteristics of the data-set used. In Sec.~\ref{sec:chemo-dyn_cha} we characterize Sculptor's stellar populations. In Sec.~\ref{sec:RRotationSignal}, we present the results of the search for l.o.s. velocity gradients. In Sec.~\ref{Discussion} we discuss the implications and possible explanations of our results in the context of all the findings in the literature. Finally, in Sec.~\ref{Conclusions} we summarize our the work, extract the main conclusions, and discuss future prospects.

\section{Data} \label{sec:Data}
In this work we make use of the $v_{los}$ and metallicity ([Fe/H]) measurements of individual stars presented in \cite{Tolstoy2023}. We refer the reader to this article for a detailed discussion of the characteristics of the data-set, the data reduction process, and the methodology to determine $v_{los}$ and [Fe/H]. 

Here suffices to say that %\subsection{data-set} \label{sec:data-set}
the spectroscopic data-set, on which the measurements are based, was acquired with the GIRAFFE spectrograph in Medusa mode, on the FLAMES instrument at the Very Large Telescope (VLT). The grating used was LR8, covering between 820.6 and 940.0 nm, with a spectral resolving power of 6500 \citep{Pasquini2002}. This wavelength region includes the near-IR Ca~II triplet (CaT) lines, a well-known and well-tested indicator of the metallicity of red giant branch (RGB) stars also in composite stellar populations \citep{Battaglia2008a, Starkenburg2010, Carrera2013, Vasquez2015}.

The data-set consists of 67 independent observations of 44 pointings, from 10 different observing programmes acquired over the period from 2003 to 2018; these were homogeneously reduced and analyzed, using the most recent ESO pipeline, and then specific software developed over many years by M. Irwin. A zero-point calibration, based on a global shift applied to the spectra for each individual field, was applied to ensure that there were no velocity offsets between different exposures. Parallaxes and proper motions from Gaia Data Release 3 (DR3) \citep{GaiaDR3}, as well as the FLAMES $v_{los}$ determinations, were jointly used in order to effectively select stars members of Sculptor and discard contaminants. 

This resulted in a sample of 1604 individual RGB stars likely members of Sculptor for which the spectrum signal-to-noise ratio (SNR) was large enough (>5) to yield precise $v_{los}$; among these, 1339 have also reliable metallicities. Since in our analysis we will use both the metallicity and $v_{los}$ information, we will mainly focus on the second set, unless explicitly said otherwise. The spectra of the selected stars have a mean SNR of 53 with a minimum of 13. The velocities have a mean uncertainty of $\pm$0.6 km s$^{-1}$, while the metallicities have mean errors around $\pm$0.1 dex. In App.~\ref{app:uncert} we test the robustness of the velocity uncertainties provided in the data-set and propose corrections for them.

\begin{figure}
    \centering
        \includegraphics[width=1\columnwidth]{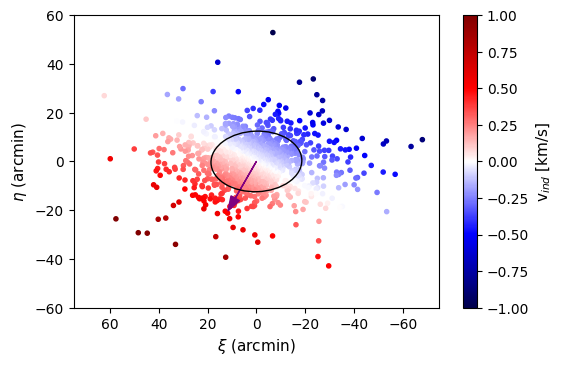}
        \caption{Projection on the sky of the spatial distribution of Sculptor's member stars, color-coded by the value of the l.o.s induced velocity because of the perspective gradient (see the color-bar). The purple arrow shows the direction of the systemic proper motion of Sculptor. The black ellipse has the ellipticity and position angle in Tab.~\ref{tab:main_char} and semi-major axis equal to twice the semi-major projected half-light radius.} 
        \label{fig:Ind_vgrad}
    \end{figure}

The large angular size of Sculptor causes the 3D systemic velocity of the galaxy to project differently over the lines-of-sight to different stars \citep{kaplinghat2008, Walker2008}. This gives rise to a perspective velocity gradient that can be mistaken for rotation. Fortunately, one can correct the l.o.s. velocities for this effect as long as the galaxy's systemic proper motion is known. To this aim, we adopted the formulae in the Appendix of \cite{Walker2008} and the systemic proper motion for Sculptor listed in Tab.~\ref{tab:main_char}. Note that these values are already corrected for the zero-points offset measured with quasars located within 7 degrees from Sculptor \citep{Battaglia2022a}. The perspective velocity gradient is visible in Fig.~\ref{fig:Ind_vgrad}: it amounts to about 1 km s$^{-1}$ in the outermost region (around 1 degree from the center) at a position angle of 37$^{\circ}$. From now on, we will work with the l.o.s. velocities corrected for the perspective velocity gradient.

\begin{table}[]
\centering

\caption{Parameters adopted for Sculptor.}
\begin{tabular}{|l|c|l|}
\hline
Parameter & Value & Ref     \\ 
\hline
    Center (RA, DEC) & (1\textsuperscript{h} 00\textsuperscript{m} 07\textsuperscript{s}, -33\textdegree\ 43\textsuperscript{'} 07\textsuperscript{"}) & 1 \\

    ($\mu_{\alpha , *}$, $\mu_{\delta}$) [$mas$ $yr^{-1}$]       & (0.098, -0.163)                & 2
\\

    Ellipticity              & 0.33               & 1           \\ 

    Position angle [º]   & 92                 & 1           \\

    Distance [kpc]       & 83.9               & 3 \\
    $R_h [arcmin]$       & 11.17               & 1 
\\

\hline

\end{tabular}

\tablefoot{Parameters used in this work. From top to bottom we list the optical center of Sculptor, its systemic proper motion, ellipticity, position angle, distance and projected half-light radius. The third column lists the corresponding references, whose numeric code corresponds to: (1) \cite{Munoz2018}; (2) \cite{Battaglia2022a}, after correcting for the zero-point offset determined with QSOs; (3) \cite{Martinezvazquez2015}. The ellipticity is $e = 1- a_1/a_2$, where $a_1$ and $a_2$ are the projected minor and major axes; the position angle is the angle of the major axis measured from north to east.}
\label{tab:main_char}
\end{table}

%--------------------------------------------------------------------

\section{Chemo-dynamical characterization} \label{sec:chemo-dyn_cha}
In Sec.~\ref{Methodology} we present the methodology used for the identification of different stellar populations. In Sec.~\ref{Result} we show the results of the chemo-dynamical characterization of Sculptor's stellar populations, and analyze the performance of the method on a mock galaxy. 

\subsection{Methodology} \label{Methodology}
According to the analysis provided by \citet{Tolstoy2023}, the existence of Sculptor's two distinct chemo-dynamical populations is not as clear as before. Rather, Sculptor seems to be possibly well described by only one population with a negative metallicity gradient. However, since the metallicity, position, and line-of-sight velocity distributions of Sculptor were analyzed separately, here we revisit the modeling jointly analyzing these quantities to properly account for possible dependencies in Sculptor's chemo-dynamical properties. In practise, we compare three different models: a single population model, where the stellar component is described by one surface number density profile and a constant l.o.s. velocity dispersion, but exhibits a metallicity gradient ({\it 1-pop}); a two-populations model, where the stellar component is described via the super-position of two components, allowing for differences in their spatial distribution, global l.o.s. velocity dispersion and metallicity distribution ({\it 2-pop}); and finally, a three-populations model ({\it 3-pop}), which we used to search for statistical evidence of some other population or residual contamination from MW stars (although it is important to point out that in our case all the stars are astrometric and l.o.s. velocity members of Sculptor, hence if present, residual contamination should be minimal).

We characterize the chemo-dynamical properties of Sculptor's stellar population broadly following the methodology presented in \citet{Walker2011, pace2020}, with some modifications. 

In Sec.~\ref{sec: Selectionfunction} we quantify the selection function of the VLT/FLAMES data-set. In Sec.~\ref{sec:likelihood} we define the likelihood function used to compare different models to the observed data. In Sec.~\ref{sec: mcmc model comparison} we explain how the parameter space is explored and how our results from different models are compared. Finally, in Sec.~\ref{sec: Probability memberships} we explain the method used to compute the probabilities of membership to each population.

\subsubsection{Selection function} \label{sec: Selectionfunction}

The number of spectroscopically observed member stars is only a subset of the overall population of Sculptor's  stars. Therefore, we include a selection function describing the ratio between spectroscopically observed member stars and overall members to Sculptor. This ratio is not uniform across the galaxy nor as a function of the stars' magnitudes, as a consequence of the placement of the fibers, of the magnitude limit and other observational conditions. 

In order to correct any selection bias, we compare our data-set with the confirmed astrometric members of \cite{Tolstoy2023}, derived using only Gaia DR3 data, considered complete down to a magnitude limit of 20 \citep{Cantat-Gaudin2023}.

Therefore, we define the selection function as: 
    \begin{equation}
        \omega(R, G) = \frac{dN_{obs}(R, G)}{dN_{mem}(R, G)},
    \end{equation}
where $dN_{obs}(R, G)$ and $dN_{mem}(R, G)$ are the number of spectroscopically observed and astrometric member stars with semi-major axis radius\footnote{defined in standard coordinates as $R = \sqrt{x^2+\left(y / (1-e)\right)^2}$, where x and y are the positions in the plane.} between $R$ and $R+dR$ and Gaia G-magnitude between $G$ and $G+dG$, respectively. 

Even though our data-set is discrete, instead of binning the data to build a discrete selection function, we adopt a smoothing Gaussian-Kernel to model it as a continuous function in the following way:
    \begin{equation} \label{eq:selecfunc}
        \omega(R, G) = \frac{{\sum_i^{N_{obs}}{\exp-\left[{\frac{\left(R_i^{obs}-R\right)^2}{k^2_R} + \frac{\left(G_i^{obs}-G\right)^2}{k^2_G}}\right]}}}{{\sum_j^{N_{mem}}{\exp-\left[{\frac{\left(R_j^{mem}-R\right)^2}{k^2_R} + \frac{\left(G_j^{mem}-G\right)^2}{k^2_G}}\right]}}},
    \end{equation}

\noindent where the subscripts $i$ and $j$ range over all the spectroscopically observed member stars and all the astrometric members, respectively. $R_i^{obs}$ ($G_i^{obs}$) and $R_j^{mem}$ ($G_j^{mem}$) are the semi-major axis radii (magnitudes) of the spectroscopically observed and astrometric member $i$ and $j$ stars, respectively; $k^2_R$ and $k^2_G$ are the smoothing parameters for the semi-major axis radius and the magnitude, respectively. The two smoothing parameters must be chosen taking into account that the density of stars in each point of the ($R$, G) plane is different; for example, the maximum density in the number of stars is found close to the center and at G$\sim$19. 

\begin{figure*}
    \centering
        \includegraphics[width=1.5\columnwidth]{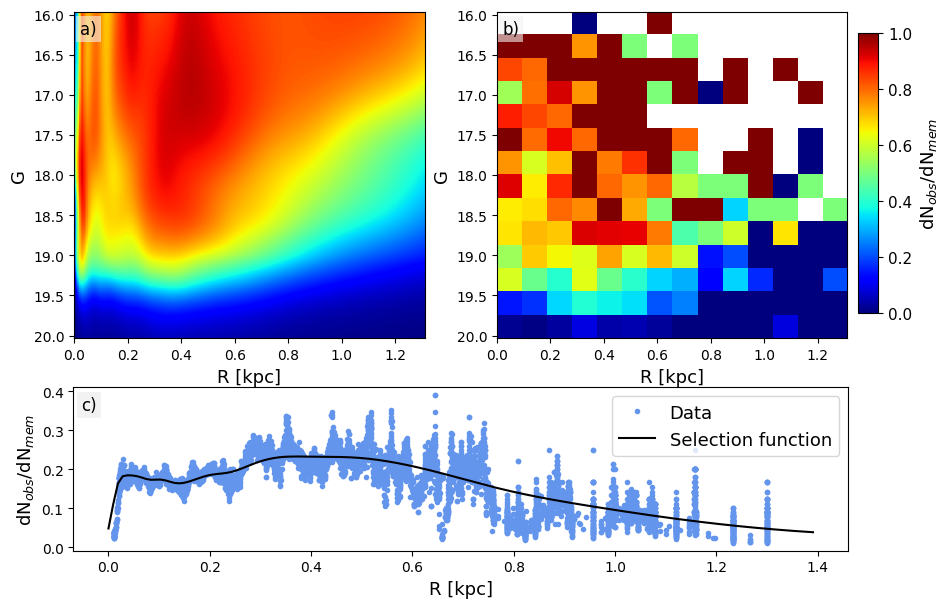}
        \caption{The selection function of the data-set. Panel a) shows the continuous selection function used in this work and defined in Eq.~\ref{eq:selecfunc}; panel b) shows the ratio between spectroscopically observed and astrometric member stars for a 15x15 gird in the ($R$, $G$) plane. Panel c) shows the selection function integrated over all magnitudes (Eq.~\ref{eq:int_selecfunc}) as a solid line and the ratio between observed and member stars for 100 different grids of binning in $R$.} 
        \label{fig:Selectionfunction}
    \end{figure*}

We adopt the following parametrizations of the softening parameters to describe our data-set:\footnote{Variations of the parametrization values for the smoothing parameters do not affect our results.} $k^2_R = 0.008 + R / 12 + R^2/10$ where $R$ is expressed in deg (so $k^2_R$ ranges from 0.5 arcmin in the inner parts to 16 arcmin in the outer part) and $k^2_G = 0.25 + \left| G-19 \right| / 6$ (from 0.25 mag in the region around G = 19 to 0.75 for the brighter magnitudes). As a comparison, panels a) and b) of Fig.~\ref{fig:Selectionfunction} compare the selection function computed using eq.~\ref{eq:selecfunc} along with the aforementioned parametrization for $k^2_R$ and $k^2_G$ (left panel), with the selection function computed binning the dataset on a 15x15 grid.

In order to simplify the terminology, we also define the integral of the selection function over all the magnitudes as: 
    \begin{equation}\label{eq:int_selecfunc}
        \Omega(R) = \int_{-\infty}^{+\infty}\omega(R, G)dG .
    \end{equation}

\noindent In panel c) of Fig.~\ref{fig:Selectionfunction} we can see $\Omega(R)$ for the chosen softening parameters, compared with the observational data for 100 different arbitrary binning grids: the general good agreement can be visually appreciated here. One can notice that there is a small discrepancy between the data and the modelled selection function in the inner parts and this might warrant a different softening parameter for those regions; however, this is difficult to determine with the data at hand, since we want to model the 2D map $G$ vs $R$, and we do not have many stars in the inner part for all the values of $G$.

\subsubsection{Likelihood} \label{sec:likelihood}
We work within the framework of Bayesian inference. Recalling Bayes theorem, we have that
    \begin{equation}
        P\left(Y\left|\boldsymbol{\zeta},M\right.\right)=\frac{P\left(\boldsymbol{\zeta}\left|Y,M\right.\right)P\left(Y\left|M\right.\right)}{P\left(\boldsymbol{\zeta}\left|M\right.\right)}.
    \end{equation}

\noindent In the above, $\boldsymbol{\zeta}$ indicates the observables that we are going to fit, $M$ the model that we are going to use to reproduce our observed data-set, and $Y$ the parameters used to describe that model.
$P\left(\boldsymbol{\zeta}\left|Y,M\right.\right)$ is the likelihood function (hereafter, $\mathcal{L}$), which represents the probability that our observational data $\boldsymbol{\zeta}$ is described by the parameters $Y$ using the model $M$. $P\left(Y\left|M\right.\right)$ is the prior probability of having a set of parameters $Y$ favored in the model $M$. $P\left(Y\left|\boldsymbol{\zeta},M\right.\right)$ is the posterior probability distribution, the probability of a set of parameters $Y$ in the model $M$ that describes our data. Finally, $P\left(\boldsymbol{\zeta}\left|M\right.\right)$ is the normalization factor.

Each star of our sample is characterized by a set of observables, $\{\boldsymbol{\zeta}_i\} = \{R_i, $[Fe/H]$_i, v_{los,i}\}$ with $i=1,....,N_{obs}$, i.e. $R_i$ the semi-major axis radius, [Fe/H]$_i$ the metallicity, $v_{los,i}$ the l.o.s. velocity of the $i$ star and $N_{obs}$ the total number of stars in the sample, respectively; when dealing with the selection function, we will also consider the star's $G$ magnitude.

For a $N_{pop}$-population model, being $N_{pop}$ the total number of stellar populations in the  model, the likelihood can be expressed as: \footnote{For details on how this likelihood is derived, see  \cite{Walker2011}.}

\begin{equation} \label{eq:likelihood}
    \log \left(\mathcal{L}\right) = 
    \sum_{i=1}^{N_{obs}}\log\left(\sum_{p=1}^{N_{pop}}\frac{f_p\,\omega(R_i, G_i)\,\mathcal{L}^p\left(\boldsymbol{\zeta}_i\right)}{\int\int\int\int  \, \omega(R, G) \, \mathcal{L}^p\left(\boldsymbol{\zeta}\right)\, d^3\boldsymbol{\zeta} dG} \right).
\end{equation}

\noindent In the above, the index $p$ runs over the populations and the index $i$ over all the stars in our sample; $f_p$ is the fraction of stars that belong to the population $p$ with respect to the total number of stars in our sample, so that $\sum_{p=1}^{Npop} f_p = 1
$; $\omega(R_i, G_i)$ is the value of the selection function corresponding to the $R$ and $G$ of the $i$ star; and finally, $\mathcal{L}^p\left(\boldsymbol{\zeta}_i\right)$ is the individual likelihood of the star $i$ belonging to the population $p$. In the denominator $d^3\boldsymbol{\zeta}$ indicates that we are integrating over $R$, [Fe/H] and $v_{los}$, and $dG$ over the  magnitude as well. As $\mathcal{L}^p$ does not depend on the magnitude $G$, we can integrate the selection function over the magnitude and the denominator reduces to $\int\int\int  \, \Omega(R) \, \mathcal{L}^p\left(\boldsymbol{\zeta}\right)\, d^3\boldsymbol{\zeta}$. 

The individual likelihood $\mathcal{L}^p$ of a star in a certain population is:

\begin{equation}
    \mathcal{L}^p\left(\boldsymbol{\zeta}_i\right) = P^p_{r}\left(R_i\right)P^p_{\mathcal{M}}\left(\left[Fe/H\right]_i,R_i\right)P^p_{\mathcal{V}}\left(v_{los,i}\right),
\end{equation}

\noindent where $P^p_{r}\left(R_i\right)$, $P^p_{\mathcal{M}}\left(\left[Fe/H\right]_i, R_i\right)$ and $P^p_{\mathcal{V}}\left(v_{los,i}\right)$ are the semi-major axis radius, metallicity and velocity probability distribution functions respectively. 
Hereafter, we will drop the super index p from the notation.

For the single population model with a metallicity gradient, \onep, the metallicity distribution depends on $R$ and only the $v_{los}$ term is separable. If we integrate over it, the denominator of eq.~\ref{eq:likelihood} becomes $\int\int \, \Omega(R) \, P_{R}\, P_{\mathcal{M}}\, dR \, d\mathcal{M}$. 

For the multiple-populations models, \twop and \threep, we assume that $P_{\mathcal{M}}$ does not depend on $R$, then $\mathcal{L}\left(\boldsymbol{\zeta}_i\right)$ is fully separable and the normalization term of the denominator in Eq.~\ref{eq:likelihood} can be directly integrated on [Fe/H] and $v_{los}$; only the radial term remains, and the denominator of Eq.~\ref{eq:likelihood} becomes $\int_0^{\infty} \, \Omega(R) \, P_{R}\, dR $.  

The stellar density profile is assumed to be a spherically symmetric Plummer distribution \citep{Plummer1911} for simplicity. Then the probability distribution of projected radii is: 
    \begin{equation}
         P_{r}\left(R_i\right) = \frac{2R_i/R_h^2}{\left(1+R_i^2/R_h^2\right)^2}, 
    \end{equation}

\noindent where $R_h$ is the projected half-light radius.

For the metallicity distribution, we assume that it follows a Gaussian:

    \begin{equation}\label{eq:metallicitydist}
        P_{\mathcal{M}}\left([Fe/H]_i\right) = \frac{1}{\sqrt{2\pi \left(\sigma_{\mathcal{M}}^2+e_{i,\mathcal{M}}^2\right)}}\,\exp\left( -\frac{1}{2}\frac{\left([Fe/H]_i-\mathcal{M}\right)^2}{\sigma_{\mathcal{M}}^2+e_{i,\mathcal{M}}^2}\right),
    \end{equation}
\noindent where, for the multiple-populations model, $\mathcal{M}$ is the mean metallicity, $\sigma_{\mathcal{M}}$ is the standard deviation of the metallicity distribution and $e_{i,\mathcal{M}}$ the errors in the $i$-th metallicity measurement. In the one-population model with the metallicity gradient, $\mathcal{M} = \mathcal{M}_0 + m\, R$, where $\mathcal{M}_0$ is the metallicity at the center of the galaxy and $m$ is the metallicity gradient.

Finally, we also assume that the $v_{los}$ follow a Gaussian distribution:

    \begin{equation}
         P_{\mathcal{V}}\left(v_{los,i}\right)=  \frac{1}{\sqrt{2\pi \left(\sigma_{\mathcal{V}}^2+e_{i,\mathcal{V}}^2\right)}}\,\exp\left( -\frac{1}{2} \frac{\left(\mathcal{V}-v_{los,i}\right)^2}{\sigma_{\mathcal{V}}^2+e_{i,\mathcal{V}}^2}\right), 
    \end{equation}
\noindent where $\mathcal{V}$ is the mean line-of-sight velocity, $\sigma_{\mathcal{V}}$ is the standard deviation of the $v_{los}$ distribution and $e_{i,\mathcal{V}}$ the errors in the l.o.s velocity measurements. 

\subsubsection{Model comparison and MCMC} \label{sec: mcmc model comparison}
 The three models that we are exploring have a different number of free parameters: \onep has six free parameters \{$R_h$, $\mathcal{M}_0$, $m$, $\sigma_{\mathcal{M}}$, $\mathcal{V}$, $\sigma_{\mathcal{V}}$\}; \twop has 11 free parameters, i.e. a set of 5 parameters (\{$R_h$, $\mathcal{M}$, $\sigma_{\mathcal{M}}$ , $\mathcal{V}$, $\sigma_{\mathcal{V}}$\}) per  population, and one extra parameter $f$ which is the fraction of stars belonging to the population 1 over the total. Finally, \threep has 17 free parameters, i.e. one set of 5 parameters per population, and two extra parameters $f_1$ and $f_2$, that are the fraction of stars belonging to population 1 and 2 over the total, respectively.
 
 To explore the parameter space and sample from the posterior distribution, we use the EMCEE package \citep{Foreman-Mackey2013}, a python implementation of the Affine Invariant Markov Chain Monte Carlo (MCMC) ensemble sampler \citep{Goodman2010}, further details about the MCMC runs are explained in App.~\ref{app:MCMCrun}. However, since the likelihoods for different models are different, and they have a different number of parameters, we cannot directly use this method to choose the best model. For that, one can use the Akaike information criteria $AIC = k-2\log(\mathcal{L})$ or the Bayesian information criteria $BIC = k\log(n)-2\log(\mathcal{L})$, where $k$ is the number of free parameters, $n$ the number of data points that we are using to compare (1339 stars in our case), and $\mathcal{L}$ the value of the median of the likelihoods for the posterior probability distribution. We use the second one, as it penalizes more strongly the extra free parameters when comparing numerous points, as in our case. A $\Delta BIC = BIC(A) - BIC(B)$ larger than 10 corresponds to very strong evidence of the A model being favored over the B one \citep{Wright2015}. Finally, to characterize the parameters that fit better our data, we define the best-fitting parameters as the median and the 16th and 18th percentiles of the posterior probability distribution for each parameter. 

\subsubsection{Probability of membership} \label{sec: Probability memberships}

The method used allows us to attribute a probability of membership of each star to a given population, we define:

\begin{equation}
    L_i^p = \frac{f_p\,\omega(R_i, G_i)\,\mathcal{L}^p\left(\boldsymbol{\zeta}_i\right)}{\int\int\int  \, \Omega(R) \, \mathcal{L}^p\left(\boldsymbol{\zeta}\right)\, d^3\boldsymbol{\zeta}}.
\end{equation}

\noindent So that, for each star $i$, the probability of belonging to the population $p$ is: 

\begin{equation} \label{eq:pmem}
    P_i^p = \frac{L_i^p}{\sum_k^{N_{pop}} L_i^k}.
\end{equation}

\noindent In our Bayesian approach, we compute the probability of membership taking into account all the posterior probability distribution. We select 5000 models according to it and compute the probability of membership for each one of them. Then we use its median, scaled so that the sum of probabilities for each star of belonging to one of the populations is 1, as the final indicator for the probability of membership.

\subsection{Results} \label{Result}
In Sec.~\ref{chemo-dyn_Sculptor} we present the results of our methodology applied to the data-set from \cite{Tolstoy2023}. As explained in App.~\ref{app:uncert}, the authors provided two data-sets in which the l.o.s. velocities and corresponding uncertainties were derived using different methods. After comparing the velocity variation with the uncertainties, we found the uncertainties not to be well estimated, and therefore, we provided some corrections. We decided to use all sets of velocities in this work, the corrected and uncorrected versions; see App.~\ref{app:uncert} for the details. Since the results produced in all the cases are the same, here we quote only those from the same velocities and uncertainties set used in \cite{Tolstoy2023}; the rest of them are shown in App.~\ref{app:Results}. In Sec.~\ref{chemo-mock} we apply the method to a mock galaxy to validate our methodology.

\subsubsection{Sculptor} \label{chemo-dyn_Sculptor}
In Tab.~\ref{tab:Quemodynfit}, we report the best-fitting parameters for the three different models, along with the $\Delta BIC$ (see Sec.~\ref{sec: mcmc model comparison}) with respect to \onep ($BIC_{1-pop} - BIC_{n-pop}$). 

In \onep, we recover the results by \cite{Tolstoy2023}, where  they found a systemic velocity of $111.2 \pm 0.2$  km s$^{-1}$  and a metallicity gradient around -0.7 dex/deg. Those values are in excellent agreement with our results, systemic velocity of $111.2 \pm 0.3$  km s$^{-1}$  and a metallicity gradient of $0.69 \pm 0.06$ dex/deg.

With a $\Delta  BIC = -40$, there is very strong evidence that \twop provides a much better description of the data than \onep. In this case, the distribution of metallicities are centered at $\mathcal{M}=-1.44^{+0.03}_{-0.03}$ for the metal-rich (hereafter, MR) and $-2.02^{+0.02}_{-0.02}$ for the metal-poor (MP) components, with a scatter of $\sigma_{\mathcal{M}} = 0.27^{+0.02}_{-0.02}$ and $\sigma_{\mathcal{M}} = 0.34^{+0.01}_{-0.01}$ respectively. The metallicity distribution function of both populations overlaps within 1$\sigma$, so they are not well differentiated when considering only this observable. The MR component has a lower l.o.s. velocity dispersion than the MP one, $\sigma_{\mathcal{V}} = 6.4^{+0.3}_{-0.4}$ km s$^{-1}$ and $11.1^{+0.3}_{-0.3}$ km s$^{-1}$,  respectively, and it is more spatially concentrated, the 2D semi-major axis half-light radii are $0.129^{+0.008}_{-0.008}$ deg and $0.27^{+0.01}_{-0.01}$ deg respectively. 

In terms of kinematic properties, we recover the results of \cite{Tolstoy2004}, \cite{Battaglia2008} and \cite{Walker2011} always within $1\sigma$, but with considerably lower errors thanks to the quality of the data-set. For the metallicity distribution we cannot directly compare with the literature as in \cite{Tolstoy2004} and \cite{Battaglia2008} the authors made a hard-cut in the metallicity to distinguish between populations and \cite{Walker2011} used the Magnesium-index rather than [Fe/H]. Regarding the extension in the sky of both components, \cite{Walker2011} found $R_h = 0.113^{+0.007}_{-0.006}$ deg for the MR and $R_h = 0.24^{+0.02}_{-0.02}$ deg for the MP using circular projected radii, that are systematically smaller than the semi-major axis projected radius. Therefore, the differences in the results are attributed to this effect. Moreover, our $R_h$ are within 1$\sigma$ compared with the ones from \cite{Battaglia2008}, obtained using the semi-major axis projected radius.

Finally, in \threep, the first two components have very similar model parameters to those already found in the two-populations case for the MR and the MP. However, the presence of a third population (pop 3, hereafeter) with a few stars is inferred\footnote{The spectra of the stars with high probability of belonging to this population have been checked to ensure that there is not any strange feature in them, such as residual cosmic rays or strong residuals from sky subtraction. Moreover, they are in general high quality spectra, with a mean/minimum SNR of 60/20.}. This population is very metal-poor with a metallicity around $\mathcal{M} = -2.90^{+0.3}_{-0.2}$; it is very extended and displaced with respect to Sculptor' center; and it displays a shift of 15 km s$^{-1}$ in the mean $v_{los}$ ($125.5^{+2.4}_{-2.6}$ km s$^{-1}$) with respect to the systemic one in Sculptor ($111.2^{+0.3}_{-0.3}$ km s$^{-1}$). This is in line with the results from \cite{Tolstoy2023}, who found an increase in the mean $v_{los}$ as a function of radius. Other works have also found extended stellar components in this dwarf galaxy \citep{Jensen2023} by analyzing the number surface density profiles derived from Gaia DR3 probable members. \threep has a $\Delta  BIC = -69$ and $\Delta  BIC = -29$ with respect to \onep and \twop, respectively. Therefore, there is statistically significant evidence to choose \threep as the model that describes best the data.

%In the literature, the authors find $R_h = 0.15 \pm 0.02$ kpc \citep{Battaglia2008} and $R_h = 0.167^{+0.010}_{-0.009}$ kpc  \citep{Walker2011} for the metal-poor and $R_h = 0.35 \pm 0.010$ kpc \citep{Battaglia2008} and $R_h = 0.33^{+0.028}_{-0.024}$ kpc \citep{Walker2011} for the metal-poor. 

\begin{table*}[]
    \centering
    
    \caption{Best fit parameters for the global properties of Sculptor's stellar component}
    \begin{tabular}{|c|c|c|c|c|c|c|c|}
    \hline
         One-population & $m$ [dex / deg] & $\mathcal{M}_0$ & $\sigma_{\mathcal{M}}$& $\mathcal{V}$ [\kms]&$\sigma_{\mathcal{V}}$ [\kms] &$R_h[deg]$& $\Delta BIC$\\
        \hline
        All & $0.69^{+0.06}_{-0.06}$ & $-1.64^{+0.02}_{-0.02}$ & $0.400^{+0.009}_{-0.008}$ & $111.2^{+0.3}_{-0.3}$ & $9.7^{+0.2}_{-0.2}$ &  $0.207^{+0.005}_{-0.005}$ &  0 \\
        \hline
        \hline
        Two-populations & $f$ & $\mathcal{M}$ & $\sigma_{\mathcal{M}}$& $\mathcal{V}$ [\kms]&$\sigma_{\mathcal{V}}$ [\kms] &$R_h[deg]$& $\Delta BIC$\\
        \hline
        MR & $0.34^{+0.04}_{-0.04}$ & $-1.44^{+0.03}_{-0.03}$ & $0.27^{+0.02}_{-0.02}$ & $111.0^{+0.4}_{-0.4}$ & $6.4^{+0.3}_{-0.3}$ &  $0.129^{+0.008}_{-0.008}$ &  $-40$ \\
        MP & $0.66^{+0.04}_{-0.04}$ & $-2.02^{+0.02}_{-0.02}$ & $0.34^{+0.01}_{-0.01}$ & $111.3^{+0.4}_{-0.4}$ & $11.1^{+0.3}_{-0.3}$ &  $0.27^{+0.01}_{-0.01}$ &   \\
        \hline
        \hline
        Three-populations & $f$ & $\mathcal{M}$ & $\sigma_{\mathcal{M}}$& $\mathcal{V}$ [\kms]&$\sigma_{\mathcal{V}}$ [\kms] & $R_h[deg]$& $\Delta BIC$\\
        \hline

        MR & $0.32^{+0.03}_{-0.03}$ & $-1.41^{+0.03}_{-0.03}$ & $0.25^{+0.02}_{-0.02}$ & $111.1^{+0.4}_{-0.4}$ & $6.5^{+0.3}_{-0.3}$ & $0.126^{+0.008}_{-0.007}$ &   \\
        MP & $0.66^{+0.03}_{-0.03}$ & $-2.00^{+0.02}_{-0.02}$ & $0.30^{+0.01}_{-0.01}$ & $110.9^{+0.4}_{-0.4}$ & $10.8^{+0.3}_{-0.3}$ & $0.26^{+0.01}_{-0.01}$  &  $-69$ \\
        Pop 3 & $0.017^{+0.007}_{-0.005}$ & $-2.9^{+0.3}_{-0.2}$ & $0.5^{+0.1}_{-0.1}$ & $125.5^{+2.4}_{-2.6}$ & $7.7^{+2.3}_{-1.7}$ & $1.32^{+0.9}_{-0.5}$ &   \\
        \hline
    \end{tabular}
    \tablefoot{Best fit parameters for the three models explored for the chemo-dynamical analysis. Col.~1 lists the population the parameter is referring to; Col.~2 shows the fraction of the total stars belonging to that population; Cols.~3 and ~4 list mean and dispersion for the metallicity distribution; Cols.~5 and ~6 list the mean and dispersion for the velocity distribution; Col.~7 show the half-light radius of the Plummer number density profile. It is important to note that since this profile does not fit properly the density distribution of the third population, its $R_h$ should not be considered representative; finally, Col.~8 lists the difference of the BIC between the one-population model and the other models.}
    \label{tab:Quemodynfit}
\end{table*}
    
\begin{figure*}
    \centering
        \includegraphics[width=1.8\columnwidth]{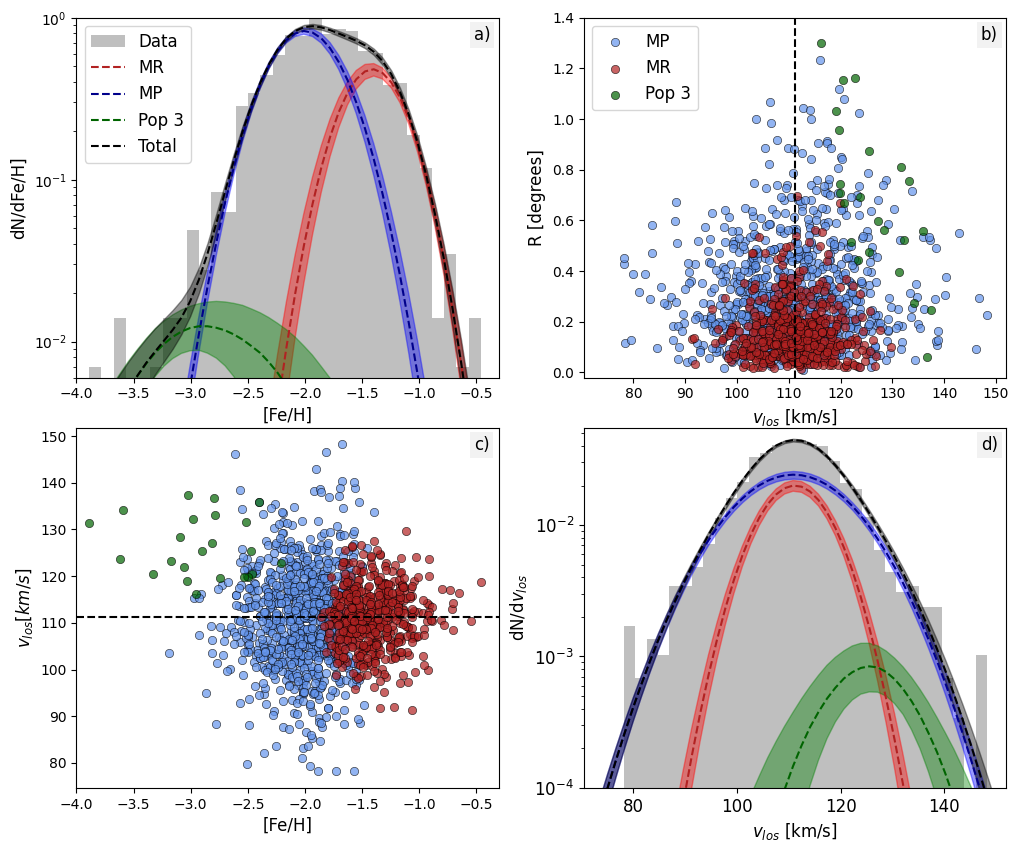}
        \caption{ a) / d) panels: Normalized metallicity/l.o.s. velocity distribution of the spectroscopically observed member stars of Sculptor fitted by the 3 populations model. The dashed-lines indicate the predicted distribution of our models: red for the MR component, blue for the MP, green for the third population, and black for the sum of the three components as indicated in the label. Bands indicate the 1$\sigma$ confidence intervals. b) / c) panels: Projection of the stars in the $v_{los}$ vs [Fe/H] / $R$ vs $v_{los}$ planes. In each panel, the stars are color-coded by the population to which they have the highest probability of belonging, red for the MR, blue for the MP and green for the third population as indicated in the label. The systemic velocity of Sculptor is indicated with straight dashed lines.}
        \label{fig:metalvelocity}
    \end{figure*}
    
In Fig.~\ref{fig:metalvelocity} we show the projection of the stars in the $v_{los}$ vs [Fe/H] and $R$ vs $v_{los}$ planes, color-coded according to the population they have the higher probability of belonging, according to Eq.~\ref{eq:pmem}, as well as the overall observed and predicted metallicity and velocity distributions. In this figure, the difference in the velocity dispersion between the MR and MP populations can be seen clearly by looking at the different spreads in the l.o.s. velocity axis. It also shows that pop 3 is more metal-poor and has a higher $v_{los}$ compared with the rest of the galaxy. Also, this component is not concentrated in the center but rather uniform in $R$. In the panel a) of Fig.~\ref{fig:metalvelocity} we show the observed metallicity distribution and the one predicted by the best-fit 3 population model. We can see that we are describing the two known populations, but also the extended metal-poor tail of the distribution with the third component. Here, we are showing the underlying  metallicity distribution obtained with the models taking into account the selection function (dashed lines and bands), and comparing it to the observed metallicity (solid histogram). We experimented with running the code without taking into account the selection function, to get an idea of how much the underlying and observed metallicity distribution differ. The result remained the same for the metallicity distribution and only the $R_h$ of the different populations changed. Therefore, we consider the one shown here to be a sufficiently good approximation of the underlined one. 

\begin{figure*}
    \centering
        \includegraphics[width=2.1\columnwidth]{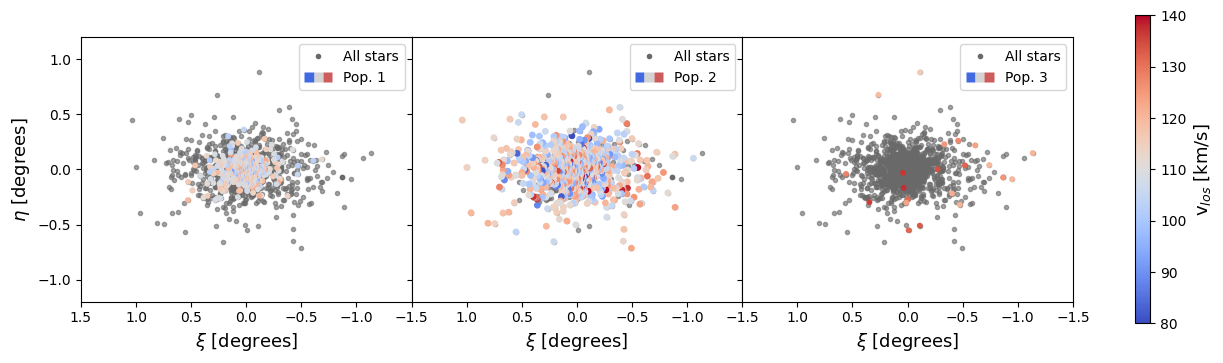}
        \caption{Projection on the sky of all stars in the LR8 sample (gray). In each panel we assign stars to each population according to the criteria explained in the text (from left to right population 1, 2 and 3) and we color-code them by $v_{los}$, as indicated in the color-bar.}
        \label{fig:Sculptorsky}
    \end{figure*}

In Fig.~\ref{fig:Sculptorsky} we show the projection of the Sculptor's stars in the sky, with the stars that belong to a specific population are color-coded by velocities. For this figure, we assigned each star to the population that yields the highest probability of membership. We can clearly see again the MR population with low velocity dispersion in the center of Sculptor and the MP population more extended and with higher velocity dispersion (left and middle panels, respectively. The third population is located in the outskirts of Sculptor, it is very extended, and its center is displaced compared to the one of Sculptor. In this figure, it is clear again that the systemic velocity of this population is shifted.

\begin{figure}
    \centering
        \includegraphics[width=\columnwidth]{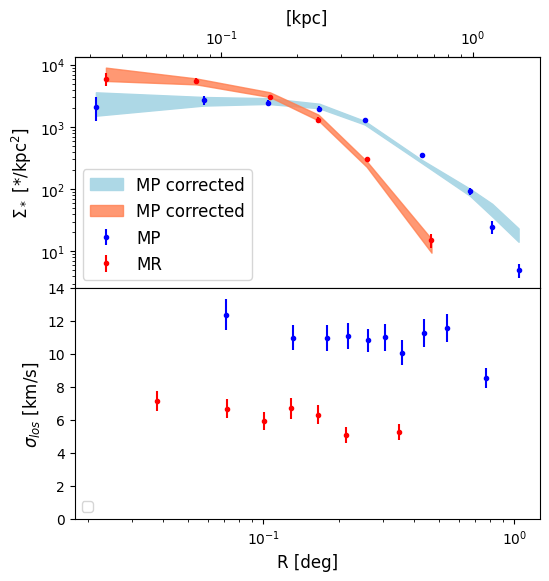}
        \caption{Upper panel: Projected surface number density profile for the MP and MR components, obtained directly by the observations and corrected by the selection function, as indicated in the legend. Lower Panel: $v_{los}$ dispersion profile for the MP and MR components, as indicated in the legend.}
        \label{fig:Profiles}
    \end{figure}

Finally, we computed the stellar surface number density and l.o.s. velocity dispersion profiles for both populations according to the membership criteria established in the previous paragraph (Fig.~\ref{fig:Profiles}). Here, stars belonging to the third population were excluded due to their distinct characteristics, and off-centered spatial distribution. The velocity dispersion, and corresponding uncertainties, were determined by fitting a Gaussian distribution to the l.o.s. velocities of the stars in each bin and accounting for velocity uncertainties \citep[see Sec~5.1 of][for details]{Taibi2018}. The number density profile has been computed by counting the number of stars in each bin, and we note that these are not the "true" underlying profiles, but rather the number of observed MR and MP stars. The selection function corrected ones are shown with bands, the main difference is that it rises the outer slope of the MP component, and it also increases slightly the density in the inner part for both components. Here we can see that the transition from the MR dominance to the MP occurs at around 0.15 deg. 

For the velocity dispersion profile, the bin size was chosen as to maintain the number of stars per bin around 80. The velocity dispersion profiles of both populations differ of $\sim$4 km s$^{-1}$ across the whole radial extend. They are slightly flatter than those obtained by \cite{Battaglia2008}. Compared with the $\sigma_{los}$ profiles computed by \cite{Strigari2017} using the data from \cite{Walker2011}, ours show more difference in velocity dispersion between both populations, mostly in the inner part. This could be because the populations are better differentiated in [Fe/H] than in the magnesium-index used in that work. These changes in the shape of the velocity dispersion profiles raise the urge to perform new dynamical modeling to infer how the dynamical mass and the dark-to-luminous mass ratio change with this new data-set. 

\subsubsection{Reliability test}  \label{chemo-mock}

In order to test the performance of our methodology, we also analyzed a mock galaxy. We chose to use one from the Gaia challenge wiki\footnote{https://astrowiki.surrey.ac.uk/doku.php?id=start} \citep{Walker2011}, so that we analyze a galaxy generated by distribution functions not related to the distributions that we used for the likelihood. This mock is a N-body realization extracted from a distribution function assuming two distinct stellar populations produced by Hernquist density profiles in a Hernquist dark matter halo. The orbits of the stars have been integrated 100 crossing times to ensure stability. For more details, we refer the reader to the documentation in the wiki, the specific file used is "SplitCompCore.data". We generate a random sample of 1339 stars of that mock galaxy. For the velocities, we use the same errors we have in our data-set for each specific stars by adding Gaussian errors to the velocities. The [Fe/H] is not provided in the mock, but the reduced magnesium index (Mg-index), and populations are not as well differentiated in Mg-index as they are in [Fe/H] for Sculptor. That is why we decided to increase the distance between the Mg-index distributions of both populations by 0.3 dex so that they are more alike to the [Fe/H] distinction we have in Sculptor. The errors are the same as the ones provided in the mock. Moreover, we added an artificial component of 25 stars resembling the third population we find in Sculptor. The parameters of the stellar populations in the mock are listed in Tab.~\ref{tab:Mock_galaxy}. The metallicities and velocities have been computed by selecting all the stars in the mock belonging to populations 1 and 2, and computing their mean value and dispersion deconvolving from errors by using Gaussian distributions. %\gbcom{This last sentence does not belong here, but earlier.} \jmcom{<- Moved upwards}. \JM{As the underlying half-light radius is not provided in the mocks, we cannot compare it.}  \gbcom{what are the spatial distributions? Did you fit them or not?}

We fit the mock using the same three different models used in the previous section. We find that \threep is the one that describes better the data, showing a $\Delta BIC$ of -322 and -122 with respect to \onep ($BIC_{1-pop} - BIC_{3-pop}$) and \twop ($BIC_{2-pop} - BIC_{3-pop}$), respectively. In Tab.~\ref{tab:Mock_galaxy} we compare the mock parameters with thones inferred from our analysis. We can clearly see that all the parameters are within 1$\sigma$ but the systemic velocity of the MR component and the velocity dispersion of the MP component which are within 2$\sigma$, this serves to validate our methodology.%\gbcom{you do not discuss the half-light radii. why?} \jmcom{corrected above}

\begin{table*}[]
    \centering
    
    \caption{Mock galaxy parameters and best fit ones}
    \begin{tabular}{|c|c|c|c|c|c|c|c|}
    \hline
        Input & $f$ & $\mathcal{M}$ & $\sigma_{\mathcal{M}}$& $\mathcal{V}$ [\kms]&$\sigma_{\mathcal{V}}$ [\kms] \\
        \hline
        MR & 0.27 & 0.61 & 0.15 & 180.4 & 16.8  \\
        MP & 0.71 & 0.12 & 0.15 & 179.7 & 22.5 \\
        Pop 3 & 0.02 & -0.5 & 0.3 & 217.8 & 10 \\
        \hline
        \hline
        Fit & $f$ & $\mathcal{M}$ & $\sigma_{\mathcal{M}}$& $\mathcal{V}$ [\kms]&$\sigma_{\mathcal{V}}$ [\kms] \\
        \hline
        MR & $0.27^{+0.02}_{-0.02}$ & $0.62^{+0.02}_{-0.02}$ & $0.15^{+0.01}_{-0.01}$ & $181.8^{+1.0}_{-1.0}$ & $17.1^{+0.8}_{-0.7}$   \\
        MP & $0.71^{+0.02}_{-0.02}$ & $0.125^{+0.009}_{-0.009}$ & $0.148^{+0.007}_{-0.007}$ & $179.3^{+0.7}_{-0.7}$& $21.7^{+0.6}_{-0.6}$\\
        Pop 3 & $0.018^{+0.004}_{-0.005}$ & $-0.5^{+0.1}_{-0.1}$ & $0.35^{+0.11}_{-0.10}$ & $219^{+3}_{-3}$& $12^{+2}_{-2}$\\
        \hline
    \end{tabular}
    \tablefoot{Comparison between mock parameters (upper part) and best fit ones after applying our chemo-dynamical analysis on a mock galaxy (lower part). Col.~1 lists the population the parameters are referred to; Cols.~2 and 3 show the mean and the width of the metallicity distributions, respectively; and, Cols.~4 and 5 list the mean and width of the velocity distributions.}
    \label{tab:Mock_galaxy}
\end{table*}

Now we use this result to analyze how well the membership probability method performs. First, we test the same criteria used for assigning the membership in Fig.~\ref{fig:Sculptorsky}, i.e. we assign a star to the population it has the highest probability of belonging. In this case, the purity\footnote{For a given population, this corresponds to the fraction of the number of true members selected over the total number of selected stars using the method} of the sample is 87, 95, 100\% for the stars assigned to populations 1, 2, 3, respectively. If, in instead of choosing the population with higher probability to belong, we put a threshold of a membership probability $>$ 80\%, the purity of the samples goes up to 95, 98, 100\% for populations 1, 2 and 3,  respectively. But at the cost of not having all stars classified into a certain population.

We performed a final reliability test, as we want to be sure that our method is not biased towards finding new populations. We run the same code on the mock galaxy but without the third population of 25 stars in the dataset. In this case, the favored model is \twop, showing a $\Delta BIC$ of -106 and -57 with respect to \onep ($BIC_{1-pop} - BIC_{2-pop}$) and \threep ($BIC_{3-pop} - BIC_{2-pop}$), respectively.

\section{Velocity gradient} \label{MRotationSignal} \label{sec:RRotationSignal}
Here we analyze the sample of 1339 probable member stars to look for signals of a velocity gradient in Sculptor. Apart from perspective effects, which we have corrected for (see Sec.~\ref{sec:Data}), a velocity gradient could be due to internal rotation or tidal effects\footnote{\cite{Martinezgarcia2023} found that a dSph would have an induced velocity gradient because of tidal effects only when approaching to the pericenter.}. However, we point out that we consider the latter hypothesis unlikely, as by re-analysis the \citet{Iorio2019} N-body simulations, tailored to reproduce a Sculptor-like galaxy on a Sculptor-like Gaia motivated orbits, show that we should not be detecting a tidally induced velocity gradient in the region we are studying. Therefore, in the following, we will use the terms "rotation" and "velocity gradient" interchangeably. 

To search for possible rotation in Sculptor, we use the same methodology as the one presented in \cite{Taibi2018}, which we summarize here. We look for evidence of rotation by comparing two different models: a radial velocity gradient model, in which apart from the velocity dispersion in the galaxy, there is a $v_{los}$ gradient signal. It is parametrized as $v_{los} = v_{sys} + nR_i\cos(\theta_i - \theta)$, where $v_{sys}$ is the systemic velocity of the galaxy, $n$ the magnitude of the gradient in km s$^{-1}$deg$^{-1}$, $R_i$ and $\theta_i$ are the distance from center of the galaxy\footnote{for the sake of completeness we decided to test this methodology using semi-major axis distance too and the results do not change significantly} and the position angle of the $i$ star, and $\theta$ is the position angle of the maximum rotation gradient, perpendicular with respect to the rotation axis;  and a dispersion-only model, in which we assume that the galaxy is fully pressure-supported. We use again a Bayesian approach to explore the posterior probability distribution, we assume Gaussian velocity distributions centered in $v_{los}$ and with a velocity dispersion $\sigma_{los}$ to build our likelihoods, so in total there are 2 free parameters for the dispersion-only model ($v_{sys}$, $\sigma_{los}$) and 4 for the rotational model ($v_{sys}$, $\sigma_{los}$, $n$, $\theta$).
For this analysis, we can compute the Bayesian evidence as the space parameter is explored using nested sampling. Therefore, in order to determine which one better describes the data, we use the Bayes factor $B_{12} = Z_1 / Z_2$, the ratio between the Bayesian evidences of two different models, 1 and 2. The Bayes factor inherently penalizes the introduction of additional free parameters in the rotational models. To compute the Bayesian evidences, we use the Multinest code \citep{Feroz2009}. The significance of one model with respect to another can be based on Jeffrey’s scale, computing the natural logarithm of the Bayes factor: values from (0-1), (1-2.5), (2.5-5), (5+) correspond to inconclusive, weak, moderate and strong evidence favoring the first model over the second one \citep{robert1995}. 

In the upper part of Tab.~\ref{tab:bestfitrotation} we report the best fit parameters of the two models. The Bayes factor between the dispersion-only model and the radial model is 0.6 favoring the radial model. However, according to Jeffry's scale, this is still inconclusive evidence of the model with rotation being favored over the dispersion-only one. In \cite{Battaglia2008} the authors found a velocity gradient along the major axis analyzing a sub-set of this FLAMES dataset. The angles of the rotation axes are different by 56 degrees, and the velocity gradient they found was -7.6$^{+3.0}_{-2.2}$ km s$^{-1}$ deg$^{-1}$, in agreement at 2$\sigma$ level with ours. It is important to note that the authors do not correct by the perspective induced velocity gradient (see Fig.~\ref{fig:Ind_vgrad}) but for the motion of the Local Standard of Rest around the Milky Way center. In \cite{Martinezgarcia2023} the authors also looked for l.o.s. velocity gradients in Sculptor using the dataset from \cite{Walker2009}, finding, however, a gradient consistent with 0 km s$^{-1}$ deg$^{-1}$ within 1$\sigma$. In this case the differences in the gradient are likely to be caused by the different datasets analyzed. 

We also look for velocity gradients independently in the MR and MP populations. We find similar results as for the general analysis: gradients of -4.4 $\pm$ 2 km s$^{-1}$ kpc$^{-1}$ and -3.0 $\pm$ 3 km s$^{-1}$ kpc$^{-1}$ with $\theta$ = 32 degrees $\pm$ 20 and $\theta$ = 25 $\pm$ 40 for the MP and MR components, respectively. However, according to the Bayes factor, the dispersion-only model is favored with inconclusive evidence for the MP and weak evidence for the MR, showing a $B_{12}$ of -0.9 and -2, respectively. So, there is no evidence of rotation in any of the populations independently, neither. This is in contrast with the results from \cite{zhu2016}, who found rotation in the MR and the MP independently around different axes. They report amplitudes of the rotational signals of 1.1 $\pm$ 0.1 km s$^{-1}$ and 0.9 $\pm$ 0.1 km s$^{-1}$ with the maximum gradient at position angles around 30 deg (changing the sign of the amplitude to match our sign) and 140 deg, for the MR and MP components respectively. The amplitudes are not directly comparable, as they assume radially constant rotation in their models, but the angles of rotation are within 1$\sigma$ for the MR component and completely off for the MP one. However, these signals could be artificially introduced by the rotational models they assume to assign the membership of the stars. This is why they test models without rotation in their App.~B, in this case, only the MR component shows rotation, which is the signal similar to the one we found. Moreover, they do not compare the results with a simpler dispersion-only model to evaluate the evidence of the signals. It is in this test of comparing models where we find that the rotation models are not a better description of the observational data for Sculptor compared to dispersion-only models.

\begin{figure}
    \centering
        \includegraphics[width=1\columnwidth]{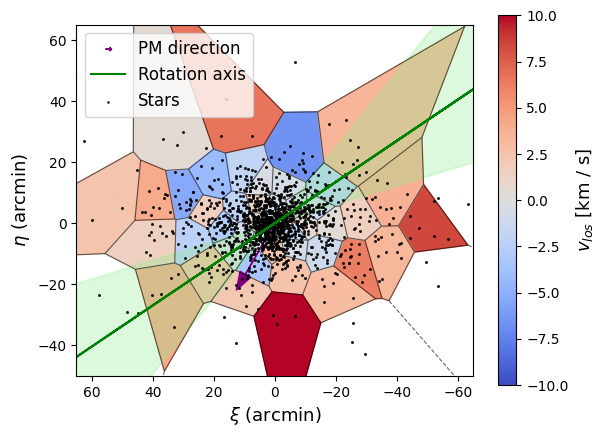}
         \includegraphics[width=0.9\columnwidth]{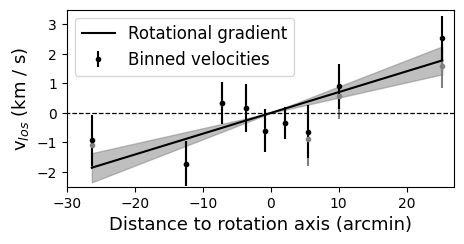}
        \caption{Upper panel: Line-of-sight velocity map of Sculptor centered in its systemic velocity and generated with Voronoi bins, it is color-coded by velocities as indicated in the color-bar. The black dots indicate the position of the stars in our sample; the purple arrow indicates the systemic proper motion of the galaxy; finally, the green line is the rotation axis, with the bands indicating the 1$\sigma$ error range. Lower panel: Velocity profile of Sculptor along the direction perpendicular to the rotation axis (maximum velocity gradient). The dots indicate the value of the mean velocity with errors in certain bins at different distances to the rotation axis, in black using all stars and in gray removing those from pop 3; the solid line shows the rotational gradient of 4.0 km s$^{-1}$ deg$^{-1}$, with the bands showing the 1$\sigma$ error range; finally, the dashed line indicates the systemic $v_{los}$ of Sculptor.}
        \label{fig:rotation}
    \end{figure}

Fig.~\ref{fig:rotation}, shows, in the top panel, the line of sight velocity field map computed using Voronoi bins with, overplotted with the rotational axis. The bottom panel shows, instead, the binned rotational gradient compared to that of the model. In the upper panel we can see a ring of low velocity stars at around 15-20 arcmins from the center, more prominent in the upper part. However, given the lack of statistics, we cannot confirm this is a physical structure. In order to see if this is an artifact of the binning procedure or not, we show velocity maps computed following different approaches in App.~\ref{app:VMAPS}. Also, in the velocity map of Fig.~\ref{fig:rotation}, there is a high-velocity region in the outer galaxy parts, more intense in the right-bottom part. This is the region where the stars belonging to the third population are located (see Fig.~\ref{fig:Sculptorsky}). To test whether those high-velocity stars are responsible for the rotation found, we repeat the same analysis, but removing the stars with larger probability of belonging to population 3 than to the MR or the MP ones. The results are listed in the lower part of Tab.~\ref{tab:bestfitrotation}. $\left<v_{los}\right>$ and $\sigma_{los}$ are very similar to the model with all the stars and within 1$\sigma$. Second, the velocity gradient found is now slightly larger but with larger uncertainties too. Third and more important, the Bayes factor between the dispersion-only model and the rotational model drops by 1.5 points to -0.9 favoring the dispersion-only model, meaning that there is even less evidence of a velocity gradient. In general, removing or not the third population, our results are in agreement with the latest results in the literature, in which they find no evidence of a $v_{los}$ gradient in Sculptor \citep{Martinezgarcia2023, Tolstoy2023}.

\begin{table}[]
    \centering
	\caption{Global kinematic parameters of Sculptor}
    \begin{tabular}{|c|c|c|c|c|c|}
    \hline
          & $v_{sys}$ & $\sigma_{los}$& $n$ &  $\theta$ & $B_{12}$\\
        
        Model & [km/s] & [km/s] & [km/s/$\circ$] &  [$\circ$] & \\
        \hline
        Disp. (2) & $111.2^{+0.2}_{-0.2}$ & $9.8^{+0.2}_{-0.2}$ & - & - & $0.6$\\
        Radial (1) & $111.2^{+0.2}_{-0.2}$ & $9.8^{+0.2}_{-0.2}$ & $-4.0^{+1.5}_{-1.5}$ & $34^{+17}_{-17}$ &  \\
        \hline

        \hline
          & $v_{los}$ & $\sigma_{los}$& $n$ &  $\theta$ & $B_{12}$\\
         Model & [km/s] & [km/s] & [km/s/$\circ$] &  [$\circ$] & \\
        \hline
        Disp. (2)& $111.0^{+0.3}_{-0.3}$ & $9.6^{+0.2}_{-0.2}$ & - & - & $-0.9$\\
        Radial (1)& $111.0^{+0.3}_{-0.3}$ & $9.6^{+0.2}_{-0.2}$ & $-4.4^{+1.7}_{-1.7}$ & $32^{+18}_{-18}$ &  \\
        \hline
    \end{tabular}
    \tablefoot {The upper part lists the results for the whole data-set, while the lower part those obtained when we remove the stars of the third population. Col.~1 lists the model the parameters are referred to; cols.~2 and 3 show the mean and dispersion of the l.o.s. velocity distribution, respectively; cols.~4 and 5 list the parameters related with the velocity gradient, i.e. its slope and the direction of the maximum velocity variation; col.~5 shows the Bayes factor $B_{12}$ between the radial rotation model (1) and the dispersion model (2).}
    \label{tab:bestfitrotation}
\end{table}

\section{Discussion} \label{Discussion}

In this section, we discuss the results in the context of the dSphs evolution. In Sec.~\ref{sec: discuss-multiplepop}, we consider the different implications of multiple populations in the context of Sculptor SFH. In Sec.~\ref{sec: discuss-Thirdpopulation} we search for the possible origin of the new third population found in Sculptor. Finally, in Sec.~\ref{sec:disc_rotation} we discuss whether there is enough evidence of a $v_{los}$ velocity gradient and the implications it has in the evolution history of Sculptor.

\subsection{Multiple populations in the context of Sculptor star formation history } \label{sec: discuss-multiplepop}
We found that a multiple populations model describes Sculptor better than a single population model with a metallicity gradient. Then there are at least two main distinct stellar components in Sculptor, a MP extended one and a MR one more concentrated and less dynamically hot. There are two different scenarios that can lead to this system: first, the metal-rich population being formed internally after the metal-poor one, or second, by external causes such as gas infall or merger events. In the former scenario, after a strong initial star-forming epoch giving rise to the MP component, the star formation was partially quenched and the gas enriched. Then, the star formation started again in a more metal-rich environment, forming the MR component. This scenario has been shown to successfully reproduce the metallicity distribution of the Sculptor dSph, according to hydrodynamical simulations of isolated dwarf galaxies \citep{Marcolini2008,Revaz2009}. The overlap in metallicity occurs as most of the gas is retained, and the enrichment can be inhomogeneous. This process of self-enrichment would take some Gyr to occur for Sculptor \citep{Babusiaux2005,Revaz2009} and would favor the extended SFHs proposed in some works in the literature \citep{deboer2012}. The second scenario is that in-falling gas or an early merger with another dwarf galaxy produced a second burst of star formation. In this case, the process could happen faster, being able to explain the short SFHs proposed in some other works in the literature \citep{Betinelli2019, delosreyes2022}. In \cite{Benitez-llambay2016} the authors analyze simulations of dSphs with two epochs of star formation, with the second one caused by interactions with another dSph. They are able to reproduce the observed characteristics of the stellar populations of dSphs like Sculptor. However, they choose galaxies with two distinct star-forming epochs separated by at least 3 Gyr in time. Given the latest reported short SFHs of Sculptor, simulations with closer star-forming epochs in time are needed to account for the possibility of short SFHs.

One could argue that any of those scenarios would have left some signatures imprinted in elemental abundances. Sculptor is known to display a "knee" in alpha elements at a metallicity [Fe/H]$\sim -1.8$ \citep{Hill2019, delosreyes2022}; this is where we start to see a significant fraction of MR members, as shown in panel c) of Fig.~\ref{fig:metalvelocity}. This supports the idea that the MP component formed quickly, therefore presenting a constant abundance of $\alpha$-elements. Subsequently, the MR population formed in an environment progressively enriched by type Ia supernovae. This could result from any of the two scenarios we presented above.

One could hypothesize that, since they are two well differentiated populations, there should be a change in the [$\alpha$ / Fe] scatter between the region where both populations coexist (-2 < [Fe/H] < -1) and the region dominated by the MR population ([Fe/H] > -1). We computed the scatter in [Mg/Fe] at different bins of [Fe/H] for the datasets of \cite{Hill2019, delosreyes2022}. However, we can not draw any conclusion because of the large uncertainties. In the future, thanks to surveys like WEAVE and 4MOST we expect to increase the statistics, so that we can test this scenario.

\subsection{Possible origins for the third population} \label{sec: discuss-Thirdpopulation}
In the following, we examine possible origins of the third population found in Sect.~\ref{chemo-dyn_Sculptor}, with a very metal-poor mean metallicity, mean velocity shifted by 15 km s$^{-1}$ with respect to the other two populations and a position offset with respect to Sculptor's center in the sky.

\subsubsection{Contamination from disk-halo stars} 
We consider the possibility that the third, very metal-poor population could be due to contamination from MW stars, belonging to the discs or to the halo. There are several reasons to discard this possibility. If the third population is made by contaminant stars, we should find more of them with velocities lower than the systemic velocity of Sculptor. Also, if they were contaminants we would expect their spatial distribution in the sky to be uniform across the area sampled. Moreover, if we compare the characteristics of these stars and the ones that have been cataloged as contaminants by \cite{Tolstoy2023} they do not follow the same trends in [Fe/H], $v_{los}$ and $R$. 
Furthermore, according to the Besançon model \citep{Robin2003}, not a single star with that $v_{los}$ and [Fe/H] is expected in the direction of Sculptor that surpasses the probability membership criteria in terms of proper motions and parallaxes used to select the possible members. Finally, we have used the methodology presented by \cite{Battaglia2022a} to assign probabilities of membership based on Gaia photometry and astrometry, position on the sky and augmented by the use of l.o.s. velocities, and the stars of this component appear as high probability members as well. We also cross-matched with a recent catalog of Sculptor members from \cite{Jensen2023}, and all the stars in common from population 3 appear to be members in their selection too.

\subsubsection{Contamination from streams}
Approximately a hundred streams have been identified in the vicinity of the MW \citep{Mateu2023}. Distributed across all the sky, these streams can reach very low metallicity and exhibit distinctive velocities, suggesting a potential explanation for the characteristics observed in the third population. There are two known streams of stars that pass close enough to Sculptor to be considered as possible sources of contamination. First, the Cetus-Palca stream \citep{Newberg2009,Thomas2022,Yuan2022}. However, even though the position in the sky coincides, in the region around Sculptor, the $v_{los}$ of the stream is around 40 km s$^{-1}$ and the proper motion is in the same direction but considerably larger than that of Sculptor. Thus, we discard Cetus-palca as a possible source of contamination. Second, there is the Murrumbidgee-G17 stream \citep{Grillmair2017}. In this case the $v_{los}$ coincides and its proper motion is roughly comparable to that of Sculptor, but it is so narrow in the sky that it could be responsible for at most 2 stars in the third component.

\subsubsection{Tides}
Another potential explanation is tidal effects due to the pericentric passages of Sculptor around the MW. 

We analyzed the N-body simulations by \cite{Iorio2019} as these were tailored to reproduce many of Sculptor's properties (3D position, 3D motion, surface number density profile, l.o.s. velocity dispersion profile) and in a set-up that would maximize the effects of tidal disturbances, in terms of MW mass, DM density profile for the simulated dwarf galaxy and Gyr of evolution. 

The analysis suggests that tidal effects are an unlikely explanation for the origin of this population. First, the streaming motions caused by tides should appear at a completely different angle, i.e. following the direction of Sculptor's systemic proper motion. Second, tidal effects would shift the velocity just a couple of km s$^{-1}$ off the systemic velocity, at the location of the stars of the third population. Finally, the effect should be symmetric, i.e. we should find a similar group of stars in the N-E direction at the same distance from the center and with a shift similar to the one of population 3, but mirrored to negative velocities.

\subsubsection{Recent minor merger}

The final physical explanation we propose is a recent minor merger. This scenario could explain the shift in position and in the mean $v_{los}$ of this population. Some candidates for the minor merger, like ultra-faint dwarfs, can easily reach the metallicity of the third population \citep{Simon2019}. On the other hand, the metallicities of pop 3 do not appear consistent with those of a stellar cluster, as the spread in metallicity of the most metal-poor clusters \citep{Beasley2019, Martin2022} is not large enough to explain the metallicity dispersion we see in here. 

\begin{figure}
    \centering
        \includegraphics[width=1\columnwidth]{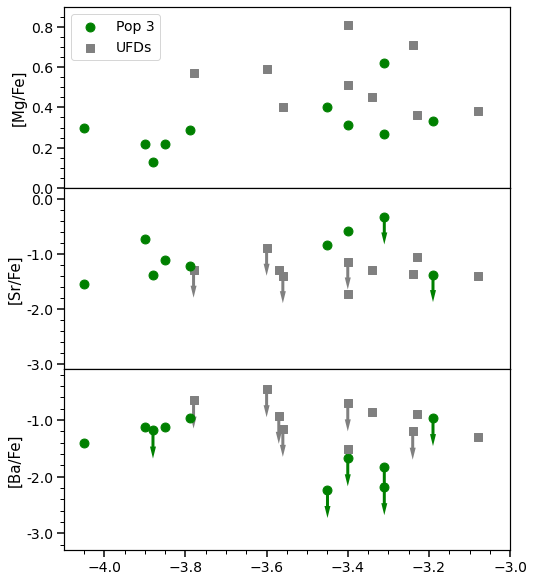}
        \caption{From top to bottom, we present the [Sr/Fe], [Ba/Fe], and [Mg/Fe] abundances as a function of [Fe/H]. The green points represent pop 3 stars with available abundances from the literature, while the gray points indicate abundances from UFDs, as indicated in the label. Downward arrows indicate upper limits. Iron abundances are from the studies in the literature.}
        \label{fig:abundances}
    \end{figure}

 In order to more directly compare the abundances of the most metal-poor stars of Sculptor and those of UFDs, and see whether pop 3 showed some specific signature, we performed a crossmatch of all our very metal-poor stars with medium-high resolution spectroscopic observations from the literature \citep{Tafelmeyer2010,Starkenburg2013, Simon2015, Jablonka2015, Skuladottir2024}, and we compared their abundances with those in UFDs \citep{Frebel2014, roederer2016,Marshall2019,Ji2019, Chiti2018,Chiti2023}. We found literature elemental abundance measurements for 10 stars belonging to the pop 3 and 5 to the MP component. They fall within the scatter and trends for UFD stars at the same metallicity, as we show in Fig.~\ref{fig:abundances}. Therefore, at present, the comparison of elemental abundances does not exclude this scenario; although it should be emphasized that the number statistics is small and scatter exists between the abundances in various UFDs. The 4MOST/4DWARFS data, with the spectra calibrated and homogenized in the same way, will be transformative by reducing the scatter in the abundance measures and increasing the statistics \citep{Skuladottirmess}.
%If we compare with other dwarf galaxies, we can see that Sculptor is slightly more Sr and Ba depleted than Fornax and Carina on average \citep{Lucchesi2024}. 

\subsubsection{Limitations of the method}
Finally, the limitations of our assumptions could be artificially producing this component. We assume that both the velocity and metallicity distributions have Gaussian shapes, however, it is common to see metal-poor tails in the metallicity distribution of dwarfs \citep[e.g.][]{Kirby2011,Leaman2012,Taibi2022}. Moreover, the models assume constant l.o.s. velocity dispersion profiles, but in Fig.~\ref{fig:Profiles} it can be clearly seen that the velocity dispersion profiles drop at least 2 km s$^{-1}$ for both, the MR and the MP populations through the radial extent. So, even though these are general assumptions and they work for distinguishing the main properties of large populations, they might not be general enough to properly characterize and differentiate such a small component as the one found here. We will address this problem by using general distribution functions to characterize the populations in Arroyo-Polonio et al. (in prep). Nonetheless, we point out that it is not a systemic difference in a single parameter what is characterizing this population. It is the combination of systematic differences in metallicities, line-of-sight velocities, and positions in the sky, what constrains statistically the existence of this component.

\subsection{Velocity gradient \label{sec:disc_rotation}}
Even though it is true that we find a model with a 3$\sigma$ detection for a velocity gradient, it does not have any statistical evidence if we compare it with a simpler dispersion-only model. Moreover, this evidence fades even more if we remove the third population of stars, whose origin is unclear. Therefore, we find no evidence of a detection of a l.o.s. velocity gradient in Sculptor.

The non-detection of a velocity gradient in Sculptor, up to this level, is also important to learn about the evolution of this galaxy. Recent literature reports that Sculptor recently passed its pericenter at around 47.7 kpc half gigayear ago \citep{Battaglia2022a} \footnote{In this work, the authors take into account both, the MW and the Large Magellanic Cloud (LMC) to compute the potential in which they integrate the orbit of Sculptor.}. Taking into account both, the MW and the LMC to integrate its orbit, it would be its first pericenter passage, considering only the MW, it would be its second one \citep{Battaglia2022a}. If we believe the scenario in which dSphs have become pressure-supported because of the interaction with the MW \citep{Mayer2001}, our results indicate that these effects have been strong enough to remove almost all the rotation in Sculptor in a single passage through the pericenter, or two if we do not consider the model with the LMC. However, the simulations from \cite{Kazantzidis2011} indicate that, for the pericentric distance of Sculptor, it is unlikely that one or two passages are enough to completely shut down the rotation of a dwarf galaxy. Moreover, other works in which they analyze the rotational state of more dwarf galaxies did not find clear evidence of a relation between the pericentric distance and the rotation support \citep{Wheeler2017, martinezgarcia2021}. Notably, even in the core Sagittarius, a highly perturbed dSph suffering the stripping of the MW, substantial rotation has been found \citep{delpino2021}, suggesting that tidal stripping alone may not be able to entirely eliminate the rotational motion.
 
\section{Conclusions}  \label{Conclusions}

In this work, we have reanalyzed the internal chemo-kinematic properties of the stellar population of the Sculptor dSph by using the new homogeneous VLT/FLAMES data-set from \cite{Tolstoy2023}, with unprecedented sample size, purity in the selection of member stars, and velocity and metallicity precision. 

We found that a model with two stellar populations with different half-light radii, mean metallicity and l.o.s. velocity dispersion describes Sculptor much better than one with a single component with a negative metallicity gradient towards the outskirts. Our best fitting models, however, includes also a third component that accounts for few very metal-poor stars, which have a clearly shifted $v_{los}$ with respect to the rest of the galaxy and a spatial distribution slightly displaced from the center. We have tested the robustness of the methodology used by applying it to mock data and found that it can correctly recover the parameters for the populations given in input with high accuracy and that it can distinguish between 2 and 3 populations systems.

In the \threep model, the MR component has a mean metallicity [Fe/H] $= -1.41^{+0.03}_{-0.03}$, half-light radius $R_h = 0.126^{+0.008}_{-0.007}$ deg and a velocity dispersion $\sigma_{los} = 6.5^{+0.3}_{-0.3}$ \kms; the MP component has a mean metallicity [Fe/H] $= -2.00^{+0.02}_{-0.02}$, half-light radius of $R_h = 0.26^{+0.01}_{-0.01}$ deg and a velocity dispersion $\sigma_{los} = 10.8^{+0.3}_{-0.3}$ \kms; the third component has a metallicity of [Fe/H] $= -2.9^{+0.3}_{-0.2}$ and a mean velocity $v_{los} = 125.5^{+2.4}_{-2.6}$ \kms. 

With this new data-set, the shape of the l.o.s. velocity dispersion for the MR and MP components that have been used in the literature for dynamical modeling of this galaxy have seen some mild variations. 

As for the group of very metal-poor stars classified in the third population, it does not seem likely that we can explain it as contamination by known MW streams, field disc or halo MW stars, and it is unlikely to be explained also by tidal effects. Therefore, the only physical explanation we find is a recent minor merger, in which the outer stars of the galaxy are not dynamically thermalized yet. Still, around 20 stars are too few to be considered as strong evidence for its presence and it cannot be excluded that limitations in the modeling assumptions are driving the detection of this component. In order to confirm the existence of this third component, more spectroscopic follow-ups on the outskirts of Sculptor will be needed.  If, with more data, we keep finding more very metal-poor stars with the offset in velocity, we could characterize this new population more accurately. 

We also revised the issue of whether Sculptor exhibits a l.o.s. velocity gradient, taking advantage of the possibility, afforded by the knowledge of Sculptor`s proper motion from Gaia eDR3 data, of correcting first for spurious velocity gradients due to perspective effects. We found inconclusive evidence of a l.o.s. velocity gradient of $-4.0^{+1.5}_{-1.5}$ km s$^{-1}$deg$^{-1}$, after correcting for the perspective gradient; this is statistically indistinguishable from a model without a l.o.s. velocity gradient. This holds both when including or when excluding stars from the third component with a shifted l.o.s. velocity distribution. Therefore this new data-set does not suggest the presence of statistically significant rotation in this galaxy.

\section*{Acknowledgements}

J. M. Arroyo acknowledges support from the Agencia Estatal de Investigación del Ministerio de Ciencia en Innovación (AEI-MICIN) and the European Social Fund (ESF+) under grant PRE2021-100638.

J. M. Arroyo, G. Battaglia, G. Thomas acknowledge support from the Agencia Estatal de Investigación del Ministerio de Ciencia en Innovación (AEI-MICIN) and the European Regional Development Fund (ERDF) under grant number AYA2017-89076-P, the AEI under grant number CEX2019-000920-S and the AEI-MICIN under grant number PID2020-118778GB-I00/10.13039/501100011033.

The authors acknowledge the referee for the constructive report, which enhanced the quality of the discussion in the manuscript.

\bibliographystyle{aa}
\bibliography{ref.bib}
  
\begin{appendix} 
\section{Analysis of the velocity uncertanties} \label{app:uncert}

The set of l.o.s. velocities on which the analysis in \cite{Tolstoy2023} is based were obtained via  a maximum-likelihood method (MLM) where 3 parameters ($v_{los}$, $\Delta v_{los}$, and the best-fit Gaussian width for blurring the delta function of the CaT template) are explored to obtain the most likely combination. However, the authors also determined an additional set of l.o.s. velocities through a cross-correlation (CC) with a template with a zero continuum and three Gaussians mimicking the CaT lines in absorption with the uncertainties being derived from the scatter of the velocities returned by fitting a Gaussian to each of the 3 CaT lines separately as in \cite{Battaglia2008a}.

Since these methodologies produce different uncertainties, this prompted us to test whether the choice of one set over the other could have an impact on our analysis.

From the \cite{Tolstoy2023} sample, \cite{Arroyo-Polonio2023} had already analyzed the l.o.s. velocity measurements from spectra of individual exposures with high SNR and for stars with $>$8 exposures over a range of years, for the purpose of monitoring velocity variations. For that sub-set of measurements, the authors found that the uncertainties for the MLM method were underestimated while those from the CC method were slightly overestimated. In that case, the observations compared were taken over several months to years in order to characterize the motion of the binaries. Thereby introducing some velocity variation, even though it may be small, it is still present. Here we perform a similar analysis, but comparing pairs of observations of the same stars taken on the same night and focusing on a larger regime of SNR, to cover the overall range of the full \cite{Tolstoy2023} sample (180 stars with spectra with 13 < SNR < 76). Even though we are using observations obtained on the same night, there is still jitter introduced by stellar granulation that can produce velocity variation on shorter time scales \citep[e.g.][]{Yu2018}. However, since we cannot model this velocity variation due to the randomness of the periodicity of these variation, this is essentially included within the uncertainties.

\begin{figure*}
    \centering
        \includegraphics[width=0.95\columnwidth]{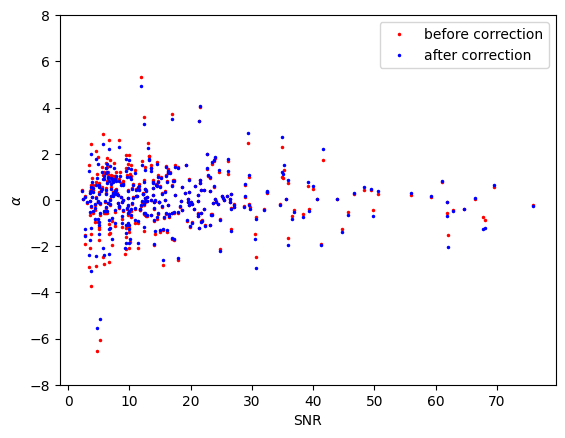}
        \includegraphics[width=0.95\columnwidth]{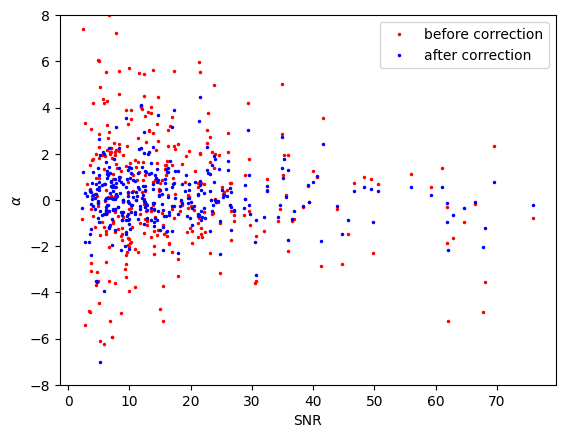}
        \includegraphics[width=0.95\columnwidth]{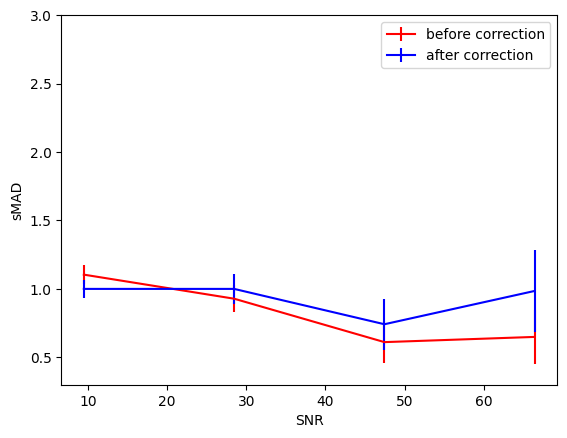}
        \includegraphics[width=0.95\columnwidth]{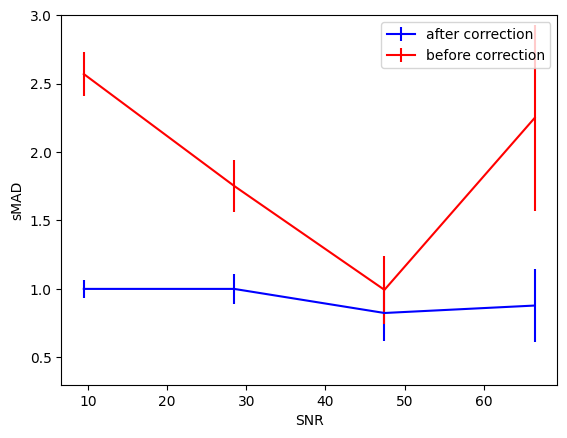}
        \caption{Upper panels: distribution of values of $\alpha$ as a function of the minimum SNR between the pair of velocities compared. In red, we show the values before the uncertainty correction and in blue after it, as indicated in the label. Left panel: CC velocities and uncertainties. Right panel: MLM velocities and uncertainties. Lower panels: sMAD of the $\alpha$ parameter distribution at different bins of SNR. In red, we show the values before the uncertainty correction and in blue after it, as indicated in the label. Left panel: CC velocities and uncertainties. Right panel: MLM velocities and uncertainties.}
        \label{Fig:alphadist}
    \end{figure*}

We compute the velocity difference normalized by the velocity uncertainties:

    \begin{equation}
        \alpha_i=\frac{v_{1}-v_{2}}{\sqrt{\left(\Delta v_{1}\right)^{2}+\left(\Delta v_{2}\right)^{2}}}
    ,\end{equation}
   
\noindent where $v_{1}$ and $v_{2}$ are the l.o.s. velocities of a given star $i$ in the two observations $1$ and $2$ in the same night, and $\Delta$ $v_{1}$ and $\Delta$ $v_{2}$ their respectively uncertainties. If the statistical uncertainties are well estimated, the scaled median absolute deviation ($sMAD$)\footnote{defined as $sMAD=1.48 \cdot$ median$|\alpha_{i}-$median$(\alpha_{i})|$} of the distribution of values of the parameter $\alpha$ should be 1. The distribution of $\alpha$ as a function of SNR ratio and the value of sMAD at different SNR are shown in red in Fig.~\ref{Fig:alphadist}.

In order to evaluate how robust our uncertainties are, we test them assuming a correction of the following form:
\begin{equation}
    \Delta v' =  \max \left[ (a - b \cdot SNR) \Delta v ,  c  \right], 
\end{equation}
so that the new uncertainties $\Delta v'$ are proportional to the old ones $\Delta v$ but corrected by a factor that scales with the signal to noise ratio ($a - b \cdot SNR$). The parameter $c$ is a minimum uncertainty floor for the highest SNRs. We divide our sample into four different regimes of SNR delimited by the values 0, 19, 38, 57 and 76. Next, we minimize the function, $f = \sum_1^4(sMAD-1)$, that characterizes how close the $sMAD$ is to 1 in each of the SNR bins, looking for the best $a, b$ and $c$ parameters. For the set of MLM velocities, we find the correction:

\begin{equation}\label{eq:mlm}
    \left(\Delta v\right)_{mlm}' =  \max \left[ (1.448 -0.021 \cdot SNR) \left(\Delta v\right)_{mlm} ,   0.65  \text{km s}^{-1} \right], 
\end{equation}

\noindent and for the set of CC velocities:

\begin{equation} \label{eq:cc}
    \left(\Delta v\right)_{cc}' =  \max \left[ (1.253 -0.011 \cdot SNR) \left(\Delta v\right)_{cc}  ,   0.53 \text{km s}^{-1}  \right].
\end{equation}

Both correction methods yield remarkably similar results in terms of $sMAD$ and final uncertainties across different SNR regimes (see Fig.~\ref{Fig:alphadist}). It is important to consider that the uncertainties $\left(\Delta v\right)_{mlm}$ are considerably lower compared to $\left(\Delta v\right)_{cc}$, but the correction is larger for them.  Despite this, the corrected MLM uncertainties $\left(\Delta v\right)'_{mlm}$ are still lower compared with the corrected CC ones $\left(\Delta v\right)'_{cc}$, meaning that the MLM has more precision for $v_{los}$ determination. Regarding the floor found, we can test it if we look at our spectral resolution. With the LR8 grating, we are resolving points in the spectra within $~$7 km s$^{-1}$ (in the wavelength regime of the CaT), and each CaT line is sampled with \textasciitilde 60 points. So this means we are fitting 180 points when we use the 3 lines. Therefore, assuming that most of the information of the $v_{los}$ comes from the CaT, the minimum theoretical error given by our spectral resolution should be around 7/$\sqrt{180}$ = 0.5 km s$^{-1}$; a value consistent with the floor we are finding. The little offset in both methods can be attributed to jittering.

In the upper panels of Fig.~\ref{Fig:alphadist} we show the values of the $\alpha$ parameter for all the pairs of observations taken in the same night, for both the CC (left) and the MLM (right) sets of velocities, before and after applying the corrections in Eq. \ref{eq:cc} and \ref{eq:mlm}. In the lower panels, we show the values of the sMAD of these distributions of $\alpha$ values. In both cases, at SNR $\sim$25 for the CC and $\sim$45 for the MLM methods, the correction factor is very close to 1.  However, when we move from that specific SNR the values of sMAD before the correction start to go away from 1, indicating a bad calibration of the uncertainties. For the corrected uncertainties, we obtain a sMAD close to 1 in all the regimes of SNR in both cases, CC and MLM. Therefore, these new uncertainties reproduce the velocity variation that we see from different exposures more robustly for all the SNR regimes. Since the corrections found in here are for the individual exposures, we combined them weighting by the errors to compute the averaged one.

\section{MCMC fitting}\label{app:MCMCrun}

    We use uniform priors for all the parameters, the range for each of them is shown in Tab.~\ref{tab:Priors}. For the multiple component populations, the priors listed are the ones we used for the first run. This serves to find the main characteristics of each population, however the walkers mix between the different populations in the chains, that is why we had to do a second run restricting the range of metallicities to those we obtained to be the likely ones for each of the components after the first run. Also for the third component the prior for the mean metallicity was changed to [-4,0].

    \begin{table}[]
    \centering
    \caption{Priors used the free parameters of each population}
    \begin{tabular}{|c|c|}
    \hline
         Parameter & Prior \\
             \hline
         $m$ [dex/deg] & [-1,1]  \\
         $\mathcal{M}_0$ & [-3,0]  \\
         $\mathcal{M}$ & [-3,0]  \\
         $\sigma_{\mathcal{M}}$ & [0,5]  \\
         $\mathcal{V}$ [\kms] & [90,130]  \\
         $\sigma_{\mathcal{V}}$ [\kms] & [0,13]  \\
         $R_h$[deg]& [0, 2]  \\
         
        \hline
    \end{tabular}
    \tablefoot{Range of the uniform priors used to run MCMC. Col.~1 lists the parameters and Col.~2 the range for the uniform prior. For populations with a metallicity gradient $\mathcal{M}_0$ and $m$ are used, while for the populations without the metallicity gradient only $\mathcal{M}$ is needed. }
    \label{tab:Priors}
\end{table}

    We sample the posterior probability distribution with a number of chains equal to 2 times the number of free parameters, each of them evolved 10000 steps. We burn 2000 of those steps, ensuring model convergence. We also applied a thinin of 32 of the order of the autocorrelation length for the chains.

\section{Results with corrected uncertainties} \label{app:Results}

In Tabs.~\ref{tab:fitmetalgradient}, ~\ref{tab:fittwopopu} and ~\ref{tab:fitthreepopu} we show the results of the chemo-dynamical analysis applied to the different data-sets: uncorrected CC, corrected CC and corrected MLM after applying the uncertainty corrections derived in  App.~\ref{app:uncert}. In all the cases the correction does not play any role in the results. This is probably because the process of averaging $v_{los}$ over different exposures for the same star mitigates its effect.

\begin{table*}[]
    \centering
    
	\caption{Best fit parameters for the global properties of Sculptor's stellar component in the case of assuming one population and a linear metallicity gradient.}
    \begin{tabular}{|c|c|c|c|c|c|c|c|c|c|c|c|c|}
    \hline
         Set &$m$ [dex / deg] & $\mathcal{M}_0$ & $\sigma_{\mathcal{M}}$& $\mathcal{V}$ [\kms]&$\sigma_{\mathcal{V}}$ [\kms] &$R_h$[deg]& $\Delta BIC$\\
        \hline
        Uncorrected CC & $-0.69^{+0.06}_{-0.06}$ & $-1.64^{+0.02}_{-0.02}$ & $0.399^{+0.009}_{-0.008}$ & $111.1^{+0.3}_{-0.3}$ & $9.5^{+0.2}_{-0.2}$ &  $0.207^{+0.005}_{-0.005}$ &  0 \\

        Corrected CC &$-0.69^{+0.06}_{-0.06}$ & $-1.64^{+0.02}_{-0.02}$ & $0.399^{+0.008}_{-0.008}$ &  $111.1^{+0.3}_{-0.3}$ & $9.5^{+0.2}_{-0.2}$ & $0.207^{+0.005}_{-0.005}$ & 0 \\
        
        Corrected MLM &$-0.69^{+0.06}_{-0.06}$ & $-1.64^{+0.02}_{-0.02}$ & $0.400^{+0.009}_{-0.008}$ &  $111.2^{+0.3}_{-0.3}$ & $9.8^{+0.2}_{-0.2}$ & $0.207^{+0.005}_{-0.005}$ & 0 \\
        \hline
    \end{tabular}
    \tablefoot{Col.~1 indicates the set of velocities used; Col.~2 indicates the metallicity gradient; Col~3 lists the mean metallicity at the center of Sculptor; Col.~4 the metallicity dispersion; Col.~5 lists the l.o.s. systemic velocity. Col.~6 the $v_{los}$ dispersion; Col.~7 indicates the half light radius assuming a Plummer number density profile; finally, Col.~8 lists the $\Delta BIC$ value with respect the single population model (to be compared to the values in Tabs ~\ref{tab:fittwopopu} and ~\ref{tab:fitthreepopu} }
    \label{tab:fitmetalgradient}
\end{table*}

\begin{table*}[]
    \centering
    
	\caption{Best fit parameters for the global properties of Sculptor's stellar component in the case of assuming two populations.}
    \begin{tabular}{|c|c|c|c|c|c|c|c|c|c|c|c|c|}
    \hline
         Set & Pop &$f$ & $\mathcal{M}$ & $\sigma_{\mathcal{M}}$& $\mathcal{V}$ [\kms]&$\sigma_{\mathcal{V}}$ [\kms] &$R_h$[deg]& $\Delta BIC$\\
        \hline
        Uncorrected CC & MR &$0.34^{+0.04}_{-0.04}$ & $-1.44^{+0.03}_{-0.04}$ & $0.27^{+0.02}_{-0.02}$ & $110.7^{+0.4}_{-0.4}$ & $6.0^{+0.3}_{-0.4}$ &  $0.128^{+0.008}_{-0.007}$ & \\
        &MP&$0.66^{+0.04}_{-0.04}$ & $-2.02^{+0.02}_{-0.02}$ & $0.34^{+0.01}_{-0.01}$ & $111.3^{+0.4}_{-0.4}$ & $10.8^{+0.3}_{-0.3}$ &  $0.272^{+0.012}_{-0.011}$ & -40 \\
        \hline
        Corrected CC & MR &$0.34^{+0.04}_{-0.04}$ & $-1.44^{+0.03}_{-0.03}$ & $0.27^{+0.02}_{-0.02}$ & $110.7^{+0.4}_{-0.4}$ & $6.1^{+0.4}_{-0.4}$ &  $0.129^{+0.008}_{-0.008}$ & \\
        &MP&$0.66^{+0.04}_{-0.04}$ & $-2.02^{+0.02}_{-0.02}$ & $0.34^{+0.01}_{-0.01}$ & $111.3^{+0.4}_{-0.4}$ & $10.8^{+0.3}_{-0.3}$ &  $0.27^{+0.01}_{-0.01}$ &  -40\\
        \hline
        Corrected MLM & MR&$0.34^{+0.04}_{-0.04}$& $-1.44^{+0.03}_{-0.03}$& $0.26^{+0.02}_{-0.02}$ & $111.0^{+0.4}_{-0.4}$& $6.4^{+0.3}_{-0.3}$& $0.128^{+0.008}_{-0.007}$& \\
        & MP&$0.66^{+0.04}_{-0.04}$& $-2.02^{+0.02}_{-0.02}$& $0.34^{+0.01}_{-0.01}$& $111.3^{+0.4}_{-0.4}$& $11.1^{+0.3}_{-0.3}$& $0.272^{+0.012}_{-0.011}$& -40\\
        \hline
    \end{tabular}
    \tablefoot{Col.~1 indicates the set of velocities used; Col.~2 lists the population we refer to; Col.~3 indicates fraction of stars belonging to the population; Col~4 lists the mean metallicity; Col.~5 the metallicity dispersion; Col.~6 lists the l.o.s. systemic velocity. Col.~7 the $v_{los}$ dispersion; Col.~8 indicates the half light radius assuming a Plummer number density profile; finally, Col.~9 lists the $\Delta BIC$ value with respect the single population model. }
    \label{tab:fittwopopu}
\end{table*}

\begin{table*}[]
    \centering
    
	\caption{Best fit parameters for the global properties of Sculptor's stellar component in the case of assuming three populations.}
    \begin{tabular}{|c|c|c|c|c|c|c|c|c|c|c|c|c|}
    \hline
         Set & Pop &$f$  & $\mathcal{M}$ & $\sigma_{\mathcal{M}}$& $\mathcal{V}$ [\kms]&$\sigma_{\mathcal{V}}$ [\kms] &$R_h$[deg]& $\Delta BIC$\\
         \hline
         Uncorrected CC & MR & $0.32^{+0.04}_{-0.03}$ & $-1.41^{+0.03}_{-0.03}$ & $0.25^{+0.02}_{-0.02}$ & $110.9^{+0.4}_{-0.4}$ & $6.1^{+0.4}_{-0.3}$ & $0.128^{+0.007}_{-0.007}$ &   \\
        &MP & $0.66^{+0.03}_{-0.04}$ & $-2.00^{+0.02}_{-0.02}$ & $0.30^{+0.01}_{-0.01}$ & $110.8^{+0.4}_{-0.4}$ & $10.5^{+0.3}_{-0.3}$ & $0.26^{+0.01}_{-0.01}$  &  $-69$ \\
        &Pop 3 & $0.018^{+0.009}_{-0.006}$ & $-2.85^{+0.24}_{-0.23}$ & $0.53^{+0.14}_{-0.14}$ & $126^{+3}_{-3}$ & $8.7^{+2.4}_{-2.0}$ & $1.1^{+0.9}_{-0.4}$ &   \\
        \hline
        Corrected CC & MR & $0.33^{+0.04}_{-0.03}$ & $-1.41^{+0.03}_{-0.03}$ & $0.25^{+0.02}_{-0.02}$ & $110.7^{+0.4}_{-0.4}$ & $6.2^{+0.3}_{-0.4}$ & $0.127^{+0.007}_{-0.008}$ &   \\
        &MP & $0.65^{+0.03}_{-0.04}$ & $-2.00^{+0.02}_{-0.02}$ & $0.30^{+0.01}_{-0.01}$ & $110.9^{+0.4}_{-0.4}$ & $10.6^{+0.3}_{-0.3}$ & $0.262^{+0.010}_{-0.010}$  &  $-69$ \\
        &Pop 3 & $0.018^{+0.009}_{-0.006}$ & $-2.82^{+0.24}_{-0.24}$ & $0.52^{+0.12}_{-0.13}$ & $125.5^{+2.7}_{-2.9}$ & $8.4^{+2.4}_{-1.8}$ & $1.6^{+1.1}_{-0.6}$ &   \\
        \hline
        Corrected MLM & MR & $0.32^{+0.04}_{-0.03}$ & $-1.41^{+0.03}_{-0.03}$ & $0.25^{+0.02}_{-0.02}$ & $111.2^{+0.4}_{-0.4}$ & $6.5^{+0.3}_{-0.3}$ & $0.126^{+0.007}_{-0.007}$ &   \\
        &MP & $0.66^{+0.03}_{-0.04}$ & $-2.00^{+0.02}_{-0.02}$ & $0.30^{+0.01}_{-0.01}$ & $110.8^{+0.4}_{-0.4}$ & $10.8^{+0.3}_{-0.3}$ & $0.26^{+0.01}_{-0.01}$  &  $-69$ \\
        &Pop 3 & $0.017^{+0.008}_{-0.006}$ & $-2.85^{+0.26}_{-0.24}$ & $0.51^{+0.14}_{-0.14}$ & $124.8^{+2.6}_{-2.7}$ & $8.3^{+2.4}_{-1.9}$ & $1.26^{+0.9}_{-0.5}$ &   \\
        \hline
    \end{tabular}
    \tablefoot{Col.~1 indicates the set of velocities used; Col.~2 lists the population we refer to; Col.~3 indicates fraction of stars belonging to the population; Col~4 lists the mean metallicity; Col.~5 the metallicity dispersion; Col.~6 lists the l.o.s. systemic velocity. Col.~7 the $v_{los}$ dispersion; Col.~8 indicates the half light radius assuming a Plummer number density profile; finally, Col.~9 lists the $\Delta BIC$ value with respect the single population model. }
    \label{tab:fitthreepopu}
\end{table*}

\section{Velocity maps}\label{app:VMAPS}
\begin{figure*}
    \centering
        \includegraphics[width=1\textwidth]{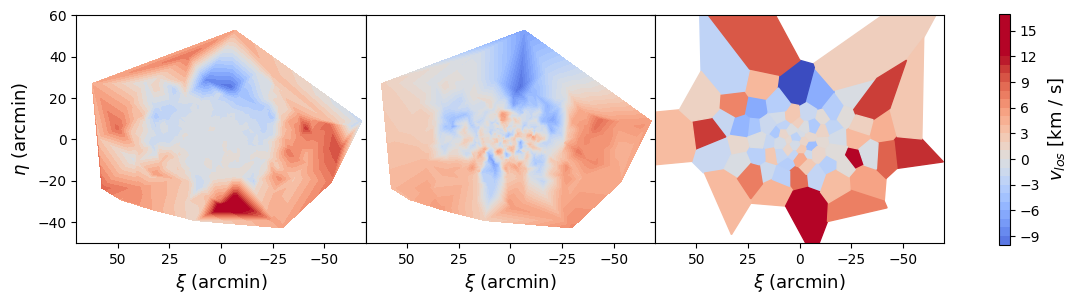}
        \caption{L.o.s. velocity maps of Sculptor. They were generated: for each point, computing the mean velocity in a radius of 12 arcmins (left); for each point, computing the mean velocity of the 25 closest neighbors (middle); binning the stars in groups using Vonoroi bins (right).}
        \label{fig:rotationx3}
    \end{figure*}

In order to characterize better whether the low velocity ring structure shown in Fig.~\ref{fig:rotation} is a physical property or whether it is an artifact caused by low-number statistics, we plotted in Fig.~\ref{fig:rotationx3} three different $v_{los}$ maps in which the velocity of each point was computed differently as indicated in the caption. In the left panel, we associate each point in the sky with the mean l.o.s. velocity of the stars closer than 12 arcmins; in the middle panel, for each point we compute the mean for the 25 closest neighbors; finally, in the right panel, we group the stars in Voronoi bins. We can see that the structure remains similar in all the cases. However, as explained in Sec.~\ref{sec:RRotationSignal}, there are not enough stars in that region to study it in detail.

\end{appendix}

\end{document}